\documentstyle[11pt,epsfig]{article}
\begin{document}

\begin{flushright}
{hep-ph/0305xxx}\\

\end{flushright}
\vspace{1cm}
\begin{center}
{\Large \bf Lepton Dipole Moments and Rare Decays in the
CP-violating MSSM with Nonuniversal Soft-Supersymmetry Breaking}
\vspace{.2cm}
\end{center}
\vspace{1cm}
\begin{center}
{Tai-Fu Feng$^{a,b,c}$\hspace{0.5cm}Tao
Huang$^{a,b}$\hspace{0.5cm} Xue-Qian Li$^{a,b,c}$\\Xin-Min
Zhang$^{a,b}$\hspace{0.5cm}Shu-Min Zhao$^{c}$\\} \vspace{.5cm}

{$^a$CCAST (World Laboratory), P.O.Box 8730,
Beijing 100080, China}\\
{$^b$Institute of High Energy Physics, Academy of Science of
China, P.O. Box
918, Beijing, 100039,  China}\\
{$^c$Department of Physics, Nankai
University, Tianjin 300070,  China}\\
\vspace{.5cm}

\end{center}
\hspace{3in}

\begin{center}
\begin{minipage}{11cm}
{\large\bf Abstract}

{\small We investigate the muon anomalous magnetic dipole moment
(MDM), the muon electric dipole moment (EDM) and the
lepton-flavour-violating decays of the $\tau-$lepton, $\tau
\rightarrow \mu \gamma$ and $\tau\rightarrow 3\mu$, in the
CP-violating Minimal Supersymmetric Standard Model (MSSM) with
nonuniversal soft-supersymmetry breaking. We evaluate numerically
the muon EDM and the branching ratios $B(\tau \rightarrow
\mu\gamma)$ and $B(\tau \rightarrow 3\mu)$, after taking into
account the experimental constraints from the electron EDM and
muon MDM. Upon imposition of the experimental limits on our
theoretical predictions for the aforementioned branching ratios
and the muon MDM, we obtain an upper bound of about
$10^{-23}\;e\cdot cm$ on the muon EDM which lies well within the
explorable reach of the proposed experiment at BNL.}

\end{minipage}
\end{center}

\vspace{4mm}
\begin{center}
\end{center}

\section{Introduction}

Recently, the Muon$-g_\mu-2$ collaboration has reported the world
average experimental value on the muon anomalous magnetic dipole
moment (MDM) \cite{Bennett}:
$$a_\mu\equiv \frac{1}{2}(g_\mu-2)=(11659203\pm 8)
\times 10^{-10},\;\;(0.7\;ppm).$$ In the framework of the standard
model (SM), the value of $a_\mu({\rm SM})$ is currently evaluated
to be \cite{Wise}
$$a_\mu({\rm SM})=(11659177\pm 7)\times 10^{-10},\;\;(0.6\;ppm).$$

The experimental value differs from the SM prediction by 1.6
standard-deviation ($1.6\sigma$). Even though the 1.6$\sigma$
deviation is not very serious, this gap might be filled up by a
contribution from new physics beyond the SM. It seems that  a
weak-scale new physics would fix the discrepancy
\cite{Komine,Mahanta}. In the framework of the SM, the
contribution to $a_\mu$ is traditionally divided into several
pieces
$$a_\mu=a_\mu^{\rm QED}+a_\mu^{\rm hadronic}+a_\mu^{\rm EW}.$$
The QED loop effects have already been computed to high
orders\cite{Kinoshita, Kinoshita1}. A thorough analysis on
hadronic contributions to the muon anomalous magnetic dipole
moment is presented in Ref.\cite{Davier}. At the one-loop level,
the contribution of the standard model is formulated as
\cite{Fujikawa,Jackiw, Altarelli, Bars, Bardeen}
$$a_{\mu}^{\rm EW}=\frac{5}{3}\frac{G_Fm_{_\mu}^2}{8\sqrt{2}\pi^2}
\bigg[1+\frac{1}{5}(1-4s_{_{\rm W}}^2)^2+{\cal O}(\frac{m_{_{\mu
}}^2}{m_{_{\rm W}}^2})\bigg].$$ The two-loop EW-sector
contributions to $a_\mu$ are also discussed in
Ref.\cite{Czarnecki}. Provided the recent measurement of $a_\mu$
is taken to be a signal of new physics beyond the SM, whose
corrections to $a_\mu$ are extensively discussed in literature,
some authors\cite{Moroi} have analyzed the muon anomalous magnetic
dipole moment in the MSSM. Considering the possible $CP$-violation
phases, Ibrahim {\it et.al} made a similar analysis in the $N=1$
supergravity model\cite{Ibrahim}. A systematic analysis on
lepton-flavor-violating processes and $a_\mu-$value within the
framework of supersymmetry see-saw mechanism was given by Hisano
{\it et.al.}\cite{Hisano}. Involving the coupling of the
second-generation with the third generation superparticles, the
contributions of R-parity violation to $a_\mu$ have been evaluated
\cite{Kim} and discussions on $a_\mu$ in the supersymmetric GUTs
are also made by some theorists \cite{Lopez}. A comparative study
on $a_\mu$ in various supersymmetric models has also been presented
in some recent works \cite{Carena}. Provided one of the neutral
Higgs bosons is of small mass, the authors of Ref.\cite{THDM}
calculated the anomalous magnetic dipole moment of muon in the
two-Higgs doublet model. An analysis of the muon anomalous
magnetic dipole moment in other extensions of the standard model
has been given \cite{new}. Alternatively, we will apply the
effective Lagrangian \cite{Arzt,Huang} to analyze the muon
anomalous magnetic dipole moment in the CP-violating MSSM with nonuniversal
soft-supersymmetry  breaking, {\it i.e.}, in this interaction
which violates the lepton flavor conservation is mediated by the
nonuniversal soft-supersymmetry breaking parameters. It is well
known that the lepton-flavor-violating decays are also ideal
processes to detect possible 'new' physics beyond the SM. So far,
the experiments have not found any substantive evidence of such
processes yet, instead, the experimental observation only sets
upper bounds on those decay branching ratios, for example
$B(\tau\rightarrow\mu\gamma)$ and $B(\tau\rightarrow 3\mu)$
\cite{PDG}. Obviously, any new physics must be constrained by
these bounds.

In this work, we investigate the lepton-flavor-violating decays
$\tau\rightarrow\mu\gamma$, $\tau\rightarrow 3\mu$, and the muon
MDM, EDM  in the framework of the MSSM with the nonuniversal
soft-supersymmetry breaking. In the supersymmetric theories, there
are many new physical $CP$-violating phases that are absent in the
SM. Considering the renormalization condition on $CP$-odd
Higgs\cite{Pilaftsis1}, we can choose the $\mu-$parameter
\footnote{Please be noted, here we use the $\mu-$parameter
following the literature, hope that it would not cause any
confusion with the muon which is sometimes written as $\mu$.} in
the superpotential, and set the non-diagonal elements of the
bilinear soft-supersymmetry breaking parameters
$m_{_{L^{IJ}}}^2,\;
m_{_{R^{IJ}}}^2,\;m_{_{Q^{IJ}}}^2,\;m_{_{U^{IJ}}}^2,\;m_{_{D^{IJ}}}^2,\;(I\neq
J)$ and soft trilinear couplings $A^u,\;A^d,\;A^l$ with the
physical $CP$ phases after properly redefining the fields  in the
theory\footnote{At the Lagrangian level, all the couplings may be
complex. However some phases are un-physical, for evaluating the
physical processes, they are not necessary and we can remove those
phases by redefining the wavefunction as $\Psi\rightarrow
e^{i\phi} \Psi$. This step is the same as to define the physical
CP phase of the CKM matrix elements in the SM case.}. Up to
one-loop order, those $CP$-violating phases induce the mixing
among the $CP$-even and $CP$-odd Higgs\cite{Pilaftsis2,Pilaftsis3}
and modify the Higgs boson couplings to the up- and down- quarks,
and to the gauge bosons drastically\cite{Pilaftsis3}. The current
experimental lower bound on the mass of the lightest neutral Higgs
can be reduced to 60 GeV \cite{Pilaftsis3}. At present, the
experimental upper bound on the electron EDM  is set\cite{PDG}:
$|d_e|<0.5\times 10^{-26}e\cdot cm$. In order to rationally
predict the  muon EDM, we need to take the electron EDM  as a
rigorous constraint into account. It is well known that the
two-loop Barr-Zee-type diagrams\cite{Barr} may also give a large
contribution to the electron EDM \cite{Pilaftsis4}, thus in our
discussion, we include the relevant two-loop Barr-Zee-type
contributions to the EDM of charged leptons .

Here, we adopt the notation of Ref.\cite{notation}, the relevant
nonuniversal soft-supersymmetry breaking terms and Feynman rules
can also be found in Ref.\cite{notation}. Our paper is organized
as follows. In Sec. II, we introduce the CP-violating MSSM with
nonuniversal soft-supersymmetry breaking. In Sec. III, we analyze
the loop-corrections to the $\bar{e}^Je^I\gamma$ effective vertex.
The muon anomalous magnetic dipole moment and the decay  width of
$\tau\rightarrow \mu\gamma$ in the supersymmetric models are
eventually formulated. The $\tau\rightarrow 3\mu$ is analyzed in
Sec.IV. Within the experimentally allowed range for the concerned
parameters, our numerical analysis is presented in Sec.V. Upon
imposition of the experimental limits on the theoretical
predictions of the branching ratios and the electron EDM and muon
MDM,  we obtain an upper bound on the muon EDM. Then we will make
a brief summary about the method and model we employ in this work
and discuss the obtained results in the last section. The tedious
formulae are collected in appendices.

\section{The MSSM with nonuniversal soft-supersymmetry breaking}

The most general form of the superpotential which has the gauge
invariance and retains all the conservation laws of the SM is
written as
\begin{eqnarray}
{\cal W}=\mu\epsilon_{ij}\hat{H}^1_i\hat{H}_j^2+\epsilon_{ij}
h_{_{IJ}}^l\hat{H}_i^1\hat{L}_j^I\hat{R}^J+\epsilon_{ij}
h_{_{IJ}}^d\hat{H}_i^1\hat{Q}_j^I\hat{D}^J+\epsilon_{ij}
h_{_{IJ}}^u\hat{H}_i^2\hat{Q}_j^I\hat{U}^J.
 \label{sp}
\end{eqnarray}
Here $\hat{H}^1,\;\hat{H}^2$ are the Higgs superfields;
$\hat{Q}^{I}$ and $\hat{L}^{I}$ are quark and lepton superfields
in doublets of the weak SU(2) group, where I=1, 2, 3 are the
indices of generations; the rest superfields $\hat{U}^{I}$,
$\hat{D}^{I}$ and $\hat{R}^{I}$ are the quark superfields of u-
and d-types and  charged leptons in singlets of the weak SU(2)
respectively. Indices i, j are contracted for the SU(2) group, and
$h^{l}$, $h^{u,d}$ are the Yukawa couplings. To break the
supersymmetry, the nonuniversal soft-supersymmetry breaking terms
are introduced as
\begin{eqnarray}
&&{\cal
L}_{soft}=-m_{_{H^1}}^2H_i^{1*}H_i^1-m_{_{H^2}}^2H_i^{2*}H_i^2
-m_{_{L^{IJ}}}^2\tilde{L}_i^{I*}\tilde{L}_i^{J}-m_{_{R^{IJ}}}^2
\tilde{R}^{I*}\tilde{R}^{J}
-m_{_{Q^{IJ}}}^2\tilde{Q}_i^{I*}\tilde{Q}_i^{J}
-m_{_{U^{IJ}}}^2\tilde{U}^{I*}\tilde{U}^{J}\nonumber \\
&&\hspace{1.4cm}-m_{_{D^{IJ}}}^2\tilde{D}^{I*}\tilde{D}^{J}
+(m_1\lambda_B\lambda_1+m_2\lambda_A^i\lambda_A^i
+m_3\lambda_G^a\lambda_G^a+h.c.) +\Big[\mu B\epsilon_{ij}H_i^1H_j^2
+\epsilon_{ij}A^l_{_{IJ}}H_{i}^{1}\tilde{L}^{I}_{j}\tilde{R}^{J}
\nonumber \\
&&\hspace{1.4cm}
+\epsilon_{ij}A^d_{_{IJ}}H_{i}^{1}\tilde{Q}^{I}_{j}\tilde{D}^{J}
+\epsilon_{ij}A^u_{_{IJ}}H_{i}^{2}\tilde{Q}^{I}_{j}
\tilde{U}^{J}+h.c.\Big],
\label{soft}
\end{eqnarray}
where
$m_{_{H^1}}^2,\;m_{_{H^2}}^2,\;m_{_{L^{IJ}}}^2,\;m_{_{R^{IJ}}}^2,
\;m_{_{Q^{IJ}}}^2,\;m_{_{U^{IJ}}}^2$ and $m_{_{D^{IJ}}}^2$ are the
square masses of the superparticles, $m_3,\;m_2,\;m_1$ denote the
masses of $\lambda_G^a\;(a=1,\;2,\;
\cdots\;8),\;\lambda_A^i\;(i=1,\;2,\;3)$ and $\lambda_B$, which
are the $SU(3)\times SU(2)\times U(1)$ gauginos. $B$ is a free
parameter in unit of mass. In ${\cal L}_{soft}$, the nonuniversal
terms are: (a): in the bilinear couplings: $
  m_{_{L^{IJ}}}^2, m_{_{U^{IJ}}}^2, m_{_{D^{IJ}}}^2$ with $I\neq J;$
whereas for I=J, $m_{_{L^{II}}}^2, m_{_{U^{II}}}^2,
m_{_{D^{II}}}^2$ are the universal soft terms; (b): in the
trilinear couplings: $  A^l_{_{IJ}}, A^u_{_{IJ}},  A^d_{_{IJ}}$
with $I \neq J$ are the nonuniversal part, whereas as I=J,
$A^l_{_{II}}, A^u_{_{II}}, A^d_{_{II}}$ are the universal part.
$A^{l}_{_{IJ}},\;A^{u}_{_{IJ}},\;A^{d}_{_{IJ}}\;(I,\;J=1,\;2,\;3)$
are the soft-supersymmetry breaking parameters that result in mass
splitting between leptons, quarks and their supersymmetric
partners. With the soft-supersymmetry breaking terms in
Eq.(\ref{soft}), we can study the phenomenology in the minimal
supersymmetric extension of the standard model (MSSM). The
resultant $6\times 6$ square-mass matrix of the charged scalar
leptons is written as
\begin{eqnarray}
&&m_{\tilde{E}}^2=\left(
\begin{array}{cc}
m_{_{L^{IJ}}}^2+(m_{e^I}^2+(\frac{1}{2}
+s_{_{\rm W}}^2)
\cos 2\beta m_{\rm Z}^2)\delta^{IJ}
& -m_{e^I}\mu^*\tan\beta\delta^{IJ}+{2m_{_{\rm W}}
s_{_{\rm W}}\over e}c_\beta A^l_{_{IJ}} \\
-m_{e^I}\mu\tan\beta\delta^{IJ}+{2m_{_{\rm W}}
s_{_{\rm W}}\over e}c_\beta A^{l*}_{_{JI}} & m_{_{R^{IJ}}}^2+(m_{e^I}^2+
s_{_{\rm W}}^2\cos 2\beta m_{\rm Z}^2)\delta^{IJ}
\end{array} \right),
\label{slmass}
\end{eqnarray}
while the $3\times 3$ sneutrino square-mass matrix is expressed as
\begin{eqnarray}
&&m_{\tilde{\nu}}^2=\left(
\begin{array}{c}-{1\over 2}\cos 2\beta m_{\rm Z}^2\delta^{IJ}+m_{_{L^{IJ}}}^2
\end{array} \right),
\label{snmass}
\end{eqnarray}
with $m_{e^I}\;(I=1,\;2,\;3)$ being the mass of the  charged
lepton of the I-th generation. Two mixing matrices
$Z_{\tilde{\nu},\tilde{E}}$ which diagonalize the square-mass
matrices of the sneutrino and charged slepton respectively are
defined as:
\begin{eqnarray}
&&{\cal Z}_{_{{\tilde \nu}}}^{\dag}m^2_{_{{\tilde \nu}}}
{\cal Z}_{_{{\tilde \nu}}}=
diag(m^2_{_{{\tilde \nu}_1}},\; m^2_{_{{\tilde \nu}_2}},
m^2_{_{{\tilde \nu}_3}})\;,
\nonumber \\
&&{\cal Z}_{_{{\tilde E}}}^{\dag}m^2_{_{{\tilde E}}}
{\cal Z}_{_{{\tilde E}}}=
diag(m^2_{_{{\tilde E}_1}},\; m^2_{_{{\tilde E}_2}},\cdots,
m^2_{_{{\tilde E}_6}})\;.
\label{zsl}
\end{eqnarray}
As for the up- and down- type scalar quarks, we can define the
mixing matrices similarly
\begin{eqnarray}
&&{\cal Z}_{_{{\tilde U}}}^{\dag}m^2_{_{{\tilde U}}}
{\cal Z}_{_{{\tilde U}}}=
diag(m^2_{_{{\tilde U}_1}},\; m^2_{_{{\tilde U}_2}},\cdots,
m^2_{_{{\tilde U}_6}})\;,\nonumber\\
&&{\cal Z}_{_{{\tilde D}}}^{\dag}m^2_{_{{\tilde D}}}
{\cal Z}_{_{{\tilde D}}}=
diag(m^2_{_{{\tilde D}_1}},\; m^2_{_{{\tilde D}_2}},\cdots,
m^2_{_{{\tilde D}_6}})\;,
\label{zsq}
\end{eqnarray}
the expressions for those mass matrices $m^2_{_{{\tilde U}}},\;m^2_{_{{\tilde D}}}$
can be found in Ref.\cite{notation}.

In order to suppress unexpectedly large effective FCNC
interactions, it is natural to assume $m_{L^{IJ}}^2\ll
m_{L^{II}}^2,\;m_{R^{IJ}}^2\ll m_{R^{II}}^2$ and $A^l_{_{IJ}}\ll
A^l_{_{II}}$ with $J\neq I$. Accurate to order ${\cal
O}(\Big({\Delta x_{_{{\tilde{S}_{ij}}}}\over
x_{_{S_i}}^2-x_{_{S_j}}^2}\Big)^2)$, we write down the expression
of the mixing matrices which diagonalize the scalar fermion
square-mass matrices
\begin{eqnarray}
&&{\cal Z}_{_{\tilde
\nu}}^{\dagger ij}={\bf U}_{_{\tilde{\nu}_{ij}}}\;,
\;\;(i,\;j=1,2,3)
\label{summ}
\end{eqnarray}
and
\begin{eqnarray}
&&{\cal Z}_{_{\tilde E}}^{\dagger \alpha i}={\bf
U}_{_{\tilde{E}_{\alpha i}}}\cos\theta_{_{\tilde{E}_i}}+{\bf
U}_{_{\tilde{E}_{\alpha(3+i)}}}\sin\theta_{_{\tilde{E}_i}}
e^{-i\varphi_{_{\tilde{E}_i}}}\;,
\nonumber\\
&&{\cal Z}_{_{\tilde E}}^{\dagger\alpha(3+i)}= -{\bf
U}_{_{\tilde{E}_{\alpha i}}}\sin\theta_{_{\tilde{E}_i}}
e^{i\varphi_{_{\tilde{E}_i}}} +{\bf
U}_{_{\tilde{E}_{\alpha(3+i)}}}\cos\theta_{_{\tilde{E}_i}}
\;\;(\alpha=1,\cdots,6;\;i=1,2,3)\;. \label{sdmm}
\end{eqnarray}

In general, the $N\times N$ transformation matrices ${\bf
U}_{_{S_{\alpha i}}}$ can be written as
\begin{eqnarray}
&&{\bf U}_{_{S_{ii}}}=1-\sum\limits_{j\neq i}
{|\Delta x_{_{S_{ij}}}|^2\over 2(x_{_{S_i}}
-x_{_{S_j}})^2}
\;,\nonumber\\
&&{\bf U}_{_{S_{ij}}}={\Delta x_{_{S_{ij}}}
\over x_{_{S_j}}
-x_{_{S_i}}}+\sum\limits_{k\neq i,\;j}
{\Delta x_{_{S_{ik}}}\Delta x_{_{S_{kj}}}\over
(x_{_{S_j}}-x_{_{S_i}})(x_{_{S_j}}
-x_{_{S_k}})}\;.
\label{dDelta}
\end{eqnarray}
where $|\Delta m^2_{_{{S_{ij}}}}|\ll
|m^2_{_{{S_{i}}}}-m^2_{_{{S_{j}}}}|$ ($i,\;j=1,\;\cdots,\;N$). The
symbols are defined as $\Delta x_{_{S_{ij}}} \equiv {\Delta
m^2_{_{{S_{ij}}}}\over m_{_{\rm W}}^2},\; x_{_{S_i}}\equiv
{m^2_{_{S_i}}\over m_{_{\rm W}}^2}$ with
$S=\tilde{\nu},\;\tilde{E}$. For the sneutrinos,  $\Delta
m^2_{_{{\tilde{\nu}_{ij}}}}(i\ne j)=m_{_{L^{ij}}}^2$. The
expressions for the off-diagonal elements of the charged slepton
square-mass matrix are more complicated:
\begin{eqnarray}
&&\Delta m_{_{\tilde{E}_{ij}}}^2=\cos\theta_{_{\tilde{E}_i}}
\cos\theta_{_{\tilde{E}_j}}m_{_{L^{ij}}}^2+\sin\theta_{_{\tilde{E}_i}}
\sin\theta_{_{\tilde{E}_j}}e^{i(\phi_{_{\tilde{E}_j}}-\phi_{_{\tilde{E}_i}}
)}m_{_{R^{ij}}}^2+{2m_{_{\rm W}}s_{_{\rm W}}c_\beta\over e}\Big[
\cos\theta_{_{\tilde{E}_j}}\sin\theta_{_{\tilde{E}_i}}
e^{-i\phi_{_{\tilde{E}_i}}}A_{ji}^{l*}\nonumber\\
&&\hspace{1.7cm}+
\cos\theta_{_{\tilde{E}_i}}\sin\theta_{_{\tilde{E}_j}}
e^{i\phi_{_{\tilde{E}_j}}}A_{ij}^{l}\Big]\;,\nonumber\\
&&\Delta m_{_{\tilde{E}_{i(3+j)}}}^2={2m_{_{\rm W}}s_{_{\rm
W}}c_\beta\over e}\Big[
\cos\theta_{_{\tilde{E}_i}}\cos\theta_{_{\tilde{E}_j}}
A_{ij}^l-\sin\theta_{_{\tilde{E}_i}}\sin\theta_{_{\tilde{E}_j}}
e^{-i(\phi_{_{\tilde{E}_i}}+\phi_{_{\tilde{E}_j}})}A_{ji}^{l*}\Big]
\nonumber\\
&&\hspace{2.2cm}-\cos\theta_{_{\tilde{E}_i}}\sin\theta_{_{\tilde{E}_j}}
e^{-i\phi_{_{\tilde{E}_j}}}m_{_{L^{ij}}}^2+\cos\theta_{_{\tilde{E}_j}}
\sin\theta_{_{\tilde{E}_i}}
e^{-i\phi_{_{\tilde{E}_i}}}m_{_{R^{ij}}}^2\;,\nonumber\\
&&\Delta m_{_{\tilde{E}_{(3+i)j}}}^2=\Delta
m_{_{\tilde{E}_{j(3+i)}}}^{2*}\;,\nonumber\\
&&\Delta
m_{_{\tilde{E}_{(3+i)(3+j)}}}^2=\cos\theta_{_{\tilde{E}_i}}
\cos\theta_{_{\tilde{E}_j}}m_{_{R^{ij}}}^2+\sin\theta_{_{\tilde{E}_i}}
\sin\theta_{_{\tilde{E}_j}}e^{i(\phi_{_{\tilde{E}_i}}-\phi_{_{\tilde{E}_j}})}
m_{_{L^{ij}}}^2\nonumber\\
&&\hspace{2.8cm}-{2m_{_{\rm W}}s_{_{\rm W}}c_\beta\over
e}\Big[\cos\theta_{_{\tilde{E}_i}}\sin\theta_{_{\tilde{E}_j}}
e^{-i\phi_{_{\tilde{E}_j}}}A_{ji}^{l*}+\cos\theta_{_{\tilde{E}_j}}
\sin\theta_{_{\tilde{E}_i}}e^{i\phi_{_{\tilde{E}_i}}}A_{ij}^l\Big]
\;,\nonumber\\
&&\Delta m_{_{\tilde{E}_{(3+i)i}}}^2=\Delta
m_{_{\tilde{E}_{i(3+i)}}}^2=0\;,\;(i,\;j=1,\;2,\;3;\;\;i\ne j).
\label{slodm}
\end{eqnarray}
For a special case where there is degeneracy among the
eigenvalues of the square-mass matrix, the explicit forms of the
mixing matrices are given in appendix \ref{app1}.

As we will find in following sections, the effective Lagrangian of
$\bar{e}^Je^I\gamma$ and $e^{^I}\rightarrow 3e^{^J}$ are mediated
by a combination of the following couplings
\begin{eqnarray}
&&{\bf G}_{_{\nu^{(a)}}}^{^{\{ijI\beta J\alpha\}}}={\cal Z}_+
^{1i*}{\cal Z}_+^{1j}{\cal Z}_{_\nu}^{I\beta*}{\cal Z}_{_\nu}^{J\alpha}
\;,\nonumber\\
&&{\bf G}_{_{\nu^{(b)}}}^{^{\{ijI\beta J\alpha\}}}={\cal Z}_+
^{1i}{\cal Z}_-^{2j}{\cal Z}_{_\nu}^{I\beta*}{\cal Z}_{_\nu}^{J\alpha}
\;,\nonumber\\
&&{\bf G}_{_{\nu^{(c)}}}^{^{\{ijI\beta J\alpha\}}}={\cal Z}_+
^{1i*}{\cal Z}_-^{2j*}{\cal Z}_{_\nu}^{I\beta*}{\cal Z}_{_\nu}^{J\alpha}
\;,\nonumber\\
&&{\bf G}_{_{\nu^{(d)}}}^{^{\{ijI\beta J\alpha\}}}={\cal Z}_-
^{2i*}{\cal Z}_-^{2j}{\cal Z}_{_\nu}^{I\beta*}{\cal Z}_{_\nu}^{J\alpha}
\;,\nonumber\\
&&{\bf G}_{_{L^{(a)}}}^{^{\{ijI\beta J\alpha\}}}=\Big[{\cal Z}_{_{\tilde{E}}}^{J\alpha*}
\Big({\cal Z}_{_N}^{1i*}s_{_{\rm W}}
+{\cal Z}_{_N}^{2i*}c_{_{\rm W}}\Big)
-{m_{_{e^J}}c_{_{\rm W}}
\over m_{_{\rm W}}c_\beta}{\cal Z}_{_{\tilde{E}}}^{(3+J)\alpha*}{\cal Z}_{_N}^{3i*}\Big]
\Big[{\cal Z}_{_{\tilde{E}}}^{I\beta}\Big({\cal Z}_{_N}^{1j}s_{_{\rm W}}
+{\cal Z}_{_N}^{2j}c_{_{\rm W}}\Big)
\nonumber\\&&\hspace{2.2cm}-{m_{_{e^I}}c_{_{\rm W}}
\over m_{_{\rm W}}c_\beta}{\cal Z}_{_{\tilde{E}}}^{(3+I)\beta}{\cal Z}_{_N}^{3j}\Big]
\;,\nonumber\\
&&{\bf G}_{_{L^{(b)}}}^{^{\{ijI\beta J\alpha\}}}=\Big[2s_{_{\rm W}}
{\cal Z}_{_{\tilde{E}}}^{(3+J)\alpha*}{\cal Z}_{_N}^{1i}+{m_{_{e^J}}c_{_{\rm W}}
\over m_{_{\rm W}}c_\beta}{\cal Z}_{_{\tilde{E}}}^{J\alpha*}{\cal Z}_{_N}^{3i}\Big]
\Big[{\cal Z}_{_{\tilde{E}}}^{I\beta}\Big({\cal Z}_{_N}^{1j}s_{_{\rm W}}
+{\cal Z}_{_N}^{2j}c_{_{\rm W}}\Big)-{m_{_{e^I}}c_{_{\rm W}}
\over m_{_{\rm W}}c_\beta}{\cal Z}_{_{\tilde{E}}}^{(3+I)\beta}{\cal Z}_{_N}^{3j}\Big]
\;,\nonumber\\
&&{\bf G}_{_{L^{(c)}}}^{^{\{ijI\beta J\alpha\}}}=\Big[{\cal Z}_{_{\tilde{E}}}^{J\alpha*}
\Big({\cal Z}_{_N}^{1i*}s_{_{\rm W}}
+{\cal Z}_{_N}^{2i*}c_{_{\rm W}}\Big)
-{m_{_{e^J}}c_{_{\rm W}}
\over m_{_{\rm W}}c_\beta}{\cal Z}_{_{\tilde{E}}}^{(3+J)\alpha*}{\cal Z}_{_N}^{3i*}\Big]
\Big[2s_{_{\rm W}}{\cal Z}_{_{\tilde{E}}}^{(3+I)\beta}{\cal Z}_{_N}^{1j*}
+{m_{_{e^I}}c_{_{\rm W}}\over m_{_{\rm W}}c_\beta}{\cal Z}_{_{\tilde{E}}}^{I\beta}
{\cal Z}_{_N}^{3j*}\Big]\;,\nonumber\\
&&{\bf G}_{_{L^{(d)}}}^{^{\{ijI\beta J\alpha\}}}=\Big[2s_{_{\rm W}}
{\cal Z}_{_{\tilde{E}}}^{(3+J)\alpha*}{\cal Z}_{_N}^{1i}+{m_{_{e^J}}c_{_{\rm W}}
\over m_{_{\rm W}}c_\beta}{\cal Z}_{_{\tilde{E}}}^{J\alpha*}{\cal Z}_{_N}^{3i}\Big]
\Big[2s_{_{\rm W}}{\cal Z}_{_{\tilde{E}}}^{(3+I)\beta}{\cal Z}_{_N}^{1j*}
+{m_{_{e^I}}c_{_{\rm W}}\over m_{_{\rm W}}c_\beta}{\cal Z}_{_{\tilde{E}}}^{I\beta}
{\cal Z}_{_N}^{3j*}\Big]\;.\label{symbol}
\end{eqnarray}
Generally, we can recast the combined couplings in terms of Eq.
(\ref{summ}) and Eq. (\ref{sdmm}). For example, we can write the
coupling ${\bf G}_{_{\nu^{(a)}}}^{^{\{ijI\beta J\alpha\}}}$ as
\begin{eqnarray}
&&{\bf G}_{_{\nu^{(a)}}}^{^{\{ijI\beta J\alpha\}}}={\cal Z}_+
^{1i*}{\cal Z}_+^{1j}\Big[\delta_{\beta I}\delta_{\alpha J}
+\delta_{\alpha J}{\Delta x_{_{\tilde{\nu}_{\beta I}}}\over
x_{_{\tilde{\nu}_I}}-x_{_{\tilde{\nu}_\beta}}}\Big|_{I\ne \beta}
+\delta_{\beta I}{\Delta x_{_{\tilde{\nu}_{\alpha J}}}^*\over
x_{_{\tilde{\nu}_J}}-x_{_{\tilde{\nu}_\alpha}}}\Big|_{J\ne \alpha}\Big]
\;,\label{exp1}
\end{eqnarray}
where the first term just contributes to the muon MDM and EDM,
while the other terms contribute to both the  muon MDM, EDM  and
the FCNC processes of leptons. The other couplings can also be
written down in a similar way, for saving space, we omit those
concrete expressions here.

It is well known that one can apply the mass insertion
approximation (MIA) to simplify the expressions of supersymmetric
contributions to flavor changing neutral current (FCNC) processes
which are induced via loop diagrams \cite{Gabbiani,Hagelin}. In
that approach, a small off-diagonal mass is inserted into the mass
matrix which is written in the basis of the weak interaction and
an approximate degeneracy of the squark masses is assumed. Thus
the drawback is obvious while evaluating some processes where
degeneracy of masses does not exist. Instead, in this work, we
carry out all the calculations rigorously in the mass basis and
keep appropriate pole masses in the propagators and mixing entries
between various flavors at the vertices. We only expand the mixing
matrix ${\cal Z}_{_{\tilde \nu}}^{\dagger ij}$ in the
soft-supersymmetry breaking terms with respect to the mixing
parameters $\Big({\Delta x_{_{{\tilde{S}_{ij}}}}\over
x_{_{S_i}}^2-x_{_{S_j}}^2}\Big)^2$ which are small as long as
$i\neq j$.

To be more explicitly,  we would like to compare our approach with
the MIA method, and point out the improvements of our scheme from
the MIA.
\begin{itemize}
\item When the mixing between left- and right-handed sfermions
is negligible, {\it i.e.} $\theta_{_{S}}\sim 0\; (S=\tilde{U}_i,\;
\tilde{D}_i,\;\tilde{E}_i)$ and $|\Delta m^2_{_{{S_{ij}}}}|\ll
|m^2_{_{{S_{i}}}}-m^2_{_{{S_{j}}}}|$ ($i,\;j=1,\;\cdots,\;N$), our
results are accord with the result of MIA at the lowest order of
$\Delta x_{_{S_{ij}}}$. When $\theta_{_{S}}\ne 0\;
(S=\tilde{U}_i,\; \tilde{D}_i,\;\tilde{E}_i)$, indeed, our
approach is an improvement from the MIA.
\item When all the eigenvalues of the mass matrix are
approximately degenerate, just as proved in \cite{Hagelin}, our
results are the same as that of MIA at the lowest order of $\Delta
x_{_{S_{ij}}}$.
\item As some eigenvalues of the mass matrix are only approximately
degenerate,  simple applications of the MIA method are invalid
\cite{Raz}. Now, there is a large flavor mixing in the sfermion
sector (certainly, such flavor changing effects will be suppressed
by the  mass of the heavy sfermion  in our detectable processes).
\end{itemize}
In our numerical results, we will extensively discuss the first
and third statements. In the following section, we give the
formulae of the muon MDM, EDM, and the decay width of
$\tau\rightarrow \mu\gamma$.

\section{The muon anomalous magnetic and electric dipole moments }

The effective Lagrangian is extensively applied to evaluating rare
decay widths of $b,\;c$-quarks\cite{Inami,Grigjanis,buras,
Grinstein}. Derivation of the Lagrangian is carried out according
to the principle: if all the masses $m_i$'s of the internal
particles in the loops  are much larger than the external momenta
i.e. $m_i^2\gg p^2$, thus the heavy particles can be integrated
out. In our case, all the SUSY particles are integrated out and
their contributions are attributed into the Wilson coefficients in
the effective Lagrangian.

For the W-boson propagator, we adopt the nonlinear $R_\xi$ gauge
whose gauge fixing term is \cite{weinberg}
\begin{equation}
{\cal L}_{_{gauge-fixing}}=-\frac{1}{\xi}f^\dagger f
\label{gaugefix}
\end{equation}
with $f=(\partial_\mu W^{+\mu}-ieA_\mu W^{+\mu}-i\xi m_{_{\rm W}}
\phi^+)$ and specifically we set $\xi=1$
in the later calculations. A thorough discussion about the gauge invariance
in this situation has been given \cite{Deshpande1,Deshpande2}.

\begin{figure}
\setlength{\unitlength}{1mm}
\begin{center}
\begin{picture}(230,200)(55,90)
\put(50,30){\includegraphics{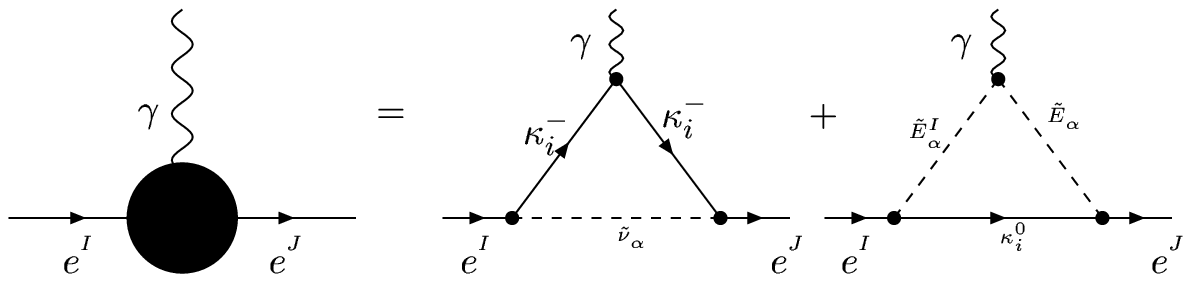}}
\end{picture}
\caption[]{The Feynman diagrams which contribute to the
$e^{^I}e^{^J}\gamma$ effective Lagrangian in  MSSM} \label{fig1}
\end{center}
\end{figure}

\begin{figure}
\setlength{\unitlength}{1mm}
\begin{center}
\begin{picture}(230,200)(55,90)
\put(50,30){\includegraphics{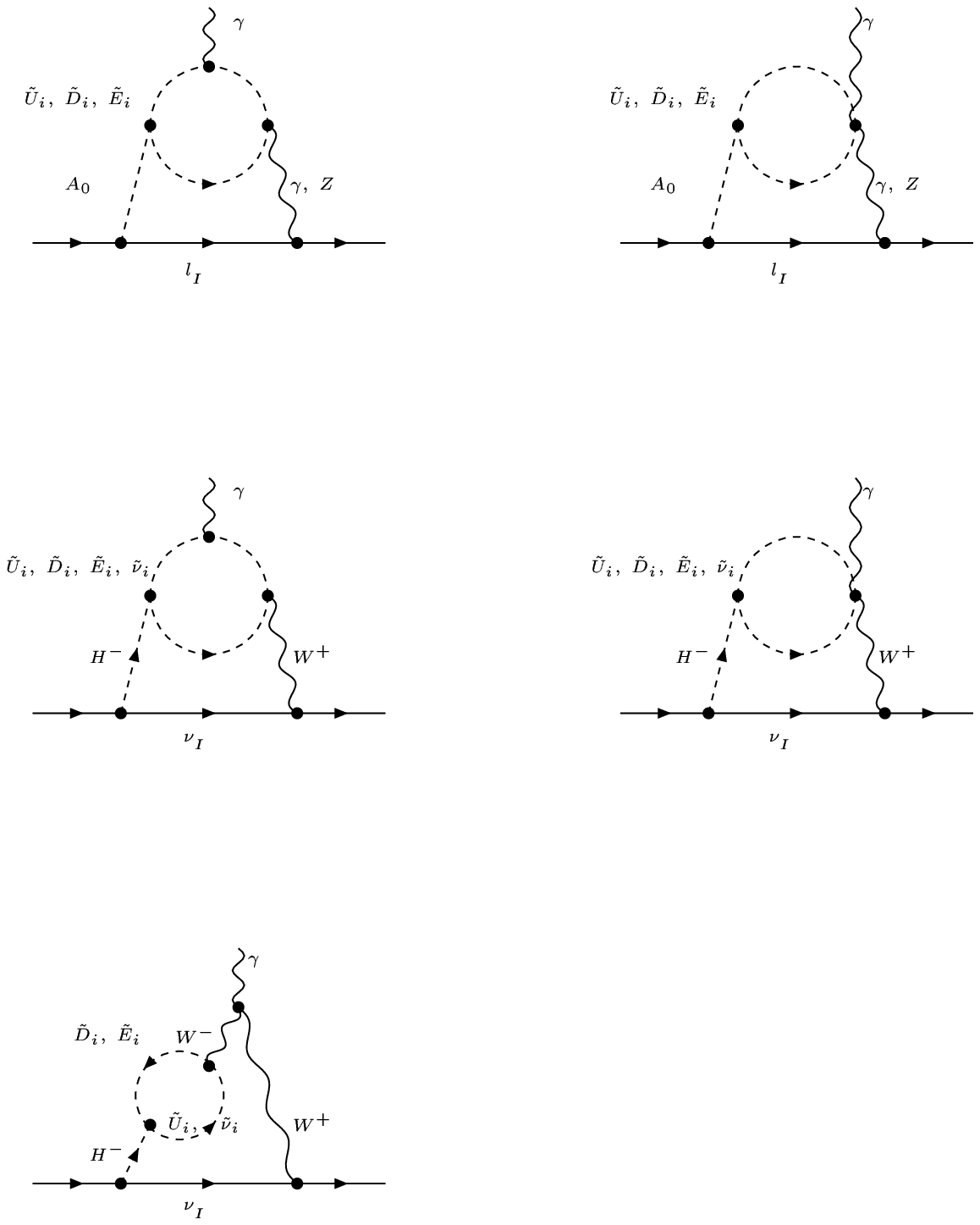}}
\end{picture}
\caption[]{The two-loop Barr-Zee-type Feynman diagrams which contribute to the
$e^{^I}e^{^J}\gamma$ effective Lagrangian in  MSSM} \label{fig2}
\end{center}
\end{figure}

The Feynman diagrams for $\bar{e}^Je^I\gamma$ in the
supersymmetric model are drawn in Fig. \ref{fig1}, the effective
Lagrangian is written as
\begin{eqnarray}
{\cal L}_{_{\bar{e}^Je^I\gamma}}={4G_F\over\sqrt{2}}
\sum\limits_{i=1}^5C_i^\mp(\mu_{_{\rm W}}){\cal O}_i^{\mp}\;,
\label{eff1}
\end{eqnarray}
and the operator basis consists of ten operators
\begin{eqnarray}
&&{\cal O}_1^\mp=\frac{1}{(4\pi)^2}\bar{e}^J(i\;/\!\!\!\!\!D)^3\omega_\mp
e^I\;,\nonumber \\
&&{\cal O}_2^\mp=\frac{1}{(4\pi)^2}\bar{e}^J\{i\;/\!\!\!\!\!D,
eF\cdot\sigma\}\omega_\mp e^I\;,
\nonumber \\
&&{\cal O}_3^\mp=\frac{1}{(4\pi)^2}\bar{e}^JiD_{\mu}(ieF^{\mu\nu})\gamma_{\nu}
\omega_\mp e^I\;,
\nonumber \\
&&{\cal O}_4^-=\frac{1}{(4\pi)^2}m_{_{e^J}}\bar{e}^J
(i\;/\!\!\!\!\!D)^2\omega_-e^I\;,
\nonumber \\
&&{\cal O}_5^-=\frac{1}{(4\pi)^2}m_{_{e^J}}
\bar{e}^JeF\cdot \sigma \omega_-e^I\;,
\nonumber \\
&&{\cal O}_4^+=\frac{1}{(4\pi)^2}m_{_{e^I}}\bar{e}^J
(i\;/\!\!\!\!\!D)^2\omega_+e^I\;,
\nonumber \\
&&{\cal O}_5^+=\frac{1}{(4\pi)^2}m_{_{e^I}}
\bar{e}^JeF\cdot \sigma \omega_+e^I\;.
\label{basmuon}
\end{eqnarray}
This basis also exists in the case of SM  \cite{Grigjanis}. In
these operators, $D_{\mu}\equiv \partial_{\mu}-ieQ_eA_{\mu}$,
$F_{\mu\nu}\equiv \partial_\mu A_\nu-\partial_\nu A_\mu$ denoting
the electromagnetic field strength tensor and $F\cdot \sigma\equiv
F_{\mu\nu}\sigma^{\mu\nu}$. The terms of dimension-four which are
related to the $\overline{e}^J\gamma_\rho\omega_\pm e^I$ vertex
cancel each other as long as we let $e^I$ and $e^J$ leptons be on
their mass shells\cite{Grigjanis}, so that they do not exist in
${\cal L}^{^{\overline{e}^Je^I\gamma}}$ at all. To shorten the
text length, we present the one-loop contributions to the Wilson
coefficients $C_i^\mp(\mu_{_{\rm W}})$ in the appendix. Here, we
give the detailed expressions for the two-loop Barr-Zee-type
diagrams (Fig.\ref{fig2}). After integrating out the heavy degrees
of freedom, we obtain the two-loop Wilson coefficients:
\begin{eqnarray}
&&C_{5(2-loop)}^+(\mu_{_{\rm W}})={e\tan\beta\over 128
\sqrt{2}\pi^2m_{_{\rm W}}s_{_{\rm W}}}\bigg\{N_c\sum\limits_{K,L,i,j}
{\cal Z}_{_{\tilde{U}}}^{Ki*}{\cal Z}_{_{\tilde{D}}}^{Lj*}
V^{KL}\Big(\Gamma^{H^+\tilde{U}_i\tilde{D}_j}\Big)^*
\Big[Q_{_u}G(x_{_{H^-}},x_{_{\tilde{U}_i}},x_{_{\tilde{D}_j}})
\nonumber\\
&&\hspace{3.0cm}
+Q_{_d}G(x_{_{H^-}},x_{_{\tilde{D}_j}},x_{_{\tilde{U}_i}})\Big]
-\sum\limits_{K,i,j}
{\cal Z}_{_{\tilde{\nu}}}^{Ki*}{\cal Z}_{_{\tilde{E}}}^{Kj*}
\Big(\Gamma^{H^+\tilde{\nu}_i\tilde{E}_j}\Big)^*
G(x_{_{H^-}},x_{_{\tilde{E}_j}},x_{_{\tilde{\nu}_i}})
\nonumber\\&&\hspace{3.0cm}+4\sqrt{2}N_c
\sum\limits_{\tilde{U},\tilde{D}}Q_{_q}G(c_{_{\rm W}}^2x_{_A}^2,
c_{_{\rm W}}^2x_{_{\tilde{Q}_i}}^2,c_{_{\rm W}}^2x_{_{\tilde{Q}_j}})
\Big[-\Gamma^{A\tilde{Q}_i^*\tilde{Q}_j}\Big(T_{_{q}}\sum\limits
_{K}{\cal Z}_{_{\tilde{Q}}}
^{Ki*}{\cal Z}_{_{\tilde{Q}}}^{Kj}-Q_{_q}^2s_{_{\rm W}}^2\delta^{ij}
\Big)\nonumber\\
&&\hspace{3.0cm}\times\Big(T_{_l}+s_{_{\rm W}}^2\Big)+
\Big(\Gamma^{A\tilde{Q}_i^*\tilde{Q}_j}\Big)^*\Big(T_{_{q}}\sum\limits
_{K}{\cal Z}_{_{\tilde{Q}}}^{Ki}{\cal Z}_{_{\tilde{Q}}}^{Kj*}
-Q_{_q}^2s_{_{\rm W}}^2\delta^{ij}\Big)s_{_{\rm W}}^2\Big]
\nonumber\\&&\hspace{3.0cm}+4\sqrt{2}
\sum\limits_{\tilde{E}}G(c_{_{\rm W}}^2x_{_A}^2,
c_{_{\rm W}}^2x_{_{\tilde{E}_i}}^2,c_{_{\rm W}}^2x_{_{\tilde{E}_j}})
\Big[\Gamma^{A\tilde{E}_i^*\tilde{E}_j}\Big(T_{_l}\sum\limits
_{K}{\cal Z}_{_{\tilde{E}}}
^{Ki*}{\cal Z}_{_{\tilde{E}}}^{Kj}-s_{_{\rm W}}^2\delta^{ij}
\Big)\nonumber\\
&&\hspace{3.0cm}\times\Big(T_{_l}+s_{_{\rm W}}^2\Big)-
\Big(\Gamma^{A\tilde{E}_i^*\tilde{E}_j}\Big)^*\Big(T_{_{l}}\sum\limits
_{K}{\cal Z}_{_{\tilde{E}}}^{Ki}{\cal Z}_{_{\tilde{E}}}^{Kj*}
-s_{_{\rm W}}^2\delta^{ij}\Big)s_{_{\rm W}}^2\Big]\nonumber\\
&&\hspace{3.0cm}
+4\sqrt{2}{s_{_{\rm W}}^2\over x_{_A}}\Big[N_c\sum\limits_{\tilde{U},\tilde{D}}
Q_q^2\Gamma^{A\tilde{Q}_j^*\tilde{Q}_j}
F({x_{_{\tilde{Q}_i}}\over x_{_A}})+\sum\limits_{\tilde{E}}
\Gamma^{A\tilde{E}_j^*\tilde{E}_j}F({x_{_{\tilde{E}_j}}\over x_{_A}})\Big]
\bigg\}\delta_{IJ}\;,\nonumber\\
&&C_{5(2-loop)}^-(\mu_{_{\rm W}})={e\tan\beta\over 128
\sqrt{2}\pi^2m_{_{\rm W}}s_{_{\rm W}}}\bigg\{N_c\sum\limits_{K,L,i,j}
{\cal Z}_{_{\tilde{U}}}^{Ki}{\cal Z}_{_{\tilde{D}}}^{Lj}
V^{KL*}\Big(\Gamma^{H^+\tilde{U}_i\tilde{D}_j}\Big)
\Big[Q_{_u}G(x_{_{H^-}},x_{_{\tilde{U}_i}},x_{_{\tilde{D}_j}})
\nonumber\\
&&\hspace{3.0cm}
+Q_{_d}G(x_{_{H^-}},x_{_{\tilde{D}_j}},x_{_{\tilde{U}_i}})\Big]
-\sum\limits_{K,i,j}{\cal Z}_{_{\tilde{\nu}}}^{Ki}
{\cal Z}_{_{\tilde{E}}}^{Kj}
\Big(\Gamma^{H^+\tilde{\nu}_i\tilde{E}_j}\Big)
G(x_{_{H^-}},x_{_{\tilde{E}_j}},x_{_{\tilde{\nu}_i}})
\nonumber\\&&\hspace{3.0cm}+4\sqrt{2}N_c
\sum\limits_{\tilde{U},\tilde{D}}Q_{_q}G(c_{_{\rm W}}^2x_{_A}^2,
c_{_{\rm W}}^2x_{_{\tilde{Q}_i}}^2,c_{_{\rm W}}^2x_{_{\tilde{Q}_j}})
\Big[-\Gamma^{A\tilde{Q}_i^*\tilde{Q}_j}\Big(T_{_{q}}\sum\limits
_{K}{\cal Z}_{_{\tilde{Q}}}
^{Ki*}{\cal Z}_{_{\tilde{Q}}}^{Kj}-Q_{_q}^2s_{_{\rm W}}^2\delta^{ij}
\Big)s_{_{\rm W}}^2\nonumber\\
&&\hspace{3.0cm}+
\Big(\Gamma^{A\tilde{Q}_i^*\tilde{Q}_j}\Big)^*\Big(T_{_{q}}\sum\limits
_{K}{\cal Z}_{_{\tilde{Q}}}^{Ki}{\cal Z}_{_{\tilde{Q}}}^{Kj*}
-Q_{_q}^2s_{_{\rm W}}^2\delta^{ij}\Big)\Big(T_{_l}+s_{_{\rm W}}^2\Big)\Big]
\nonumber\\&&\hspace{3.0cm}+4\sqrt{2}
\sum\limits_{\tilde{E}}G(c_{_{\rm W}}^2x_{_A}^2,
c_{_{\rm W}}^2x_{_{\tilde{E}_i}}^2,c_{_{\rm W}}^2x_{_{\tilde{E}_j}})
\Big[\Gamma^{A\tilde{E}_i^*\tilde{E}_j}\Big(T_{_l}\sum\limits
_{K}{\cal Z}_{_{\tilde{E}}}
^{Ki*}{\cal Z}_{_{\tilde{E}}}^{Kj}-s_{_{\rm W}}^2\delta^{ij}
\Big)s_{_{\rm W}}^2\nonumber\\
&&\hspace{3.0cm}-
\Big(\Gamma^{A\tilde{E}_i^*\tilde{E}_j}\Big)^*\Big(T_{_{l}}\sum\limits
_{K}{\cal Z}_{_{\tilde{E}}}^{Ki}{\cal Z}_{_{\tilde{E}}}^{Kj*}
-s_{_{\rm W}}^2\delta^{ij}\Big)\Big(T_{_l}+s_{_{\rm W}}^2\Big)\Big]\nonumber\\
&&\hspace{3.0cm}-4\sqrt{2}{s_{_{\rm W}}^2\over x_{_A}}
\Big[N_c\sum\limits_{\tilde{U},\tilde{D}}
Q_q^2\Gamma^{A\tilde{Q}_j^*\tilde{Q}_j}
F({x_{_{\tilde{Q}_i}}\over x_{_A}})+\sum\limits_{\tilde{E}}
\Gamma^{A\tilde{E}_j^*\tilde{E}_j}
F({x_{_{\tilde{E}_j}}\over x_{_A}})\Big]\bigg\}\delta_{IJ}
\label{2wil}
\end{eqnarray}
with $q=u_i,\;d_i,\;Q_{u_i}={2\over 3},\;Q_{d_i}={1\over 3}
,\;T_{u_i}=-T_{d_i}=-T_{l_i}={1\over 2}\;(i=1,2,3)$ and $N_c=3$ is
the color factor. The two-loop functions $F(a),\;G(a,b,c)$ are
given by
\begin{eqnarray}
&&F(a)=\int_0^1dx{x(1-x)\over a-x(1-x)}\ln\Big({x(1-x)\over a}\Big)\;,
\nonumber\\
&&G(a,b,c)=\int_0^1dx x\Big\{{ax(1-x)\ln a\over (a-1)[ax(1-x)
-bx-c(1-x)]}\nonumber\\
&&\hspace{2.0cm}+{x(1-x)[bx+c(1-x)]\over[ax(1-x)-bx-c(1-x)]
[x(1-x)-bx-c(1-x)]}\ln\Big({bx+c(1-x)\over x(1-x)}\Big)\Big\}\;.
\label{2lf}
\end{eqnarray}
The couplings between the scalar quarks and Higgs are presented in
the appendix \ref{apd1}. Integrating out the heavy degrees of
freedom in the loops,  effective vertices of $\gamma HV$ where $H$
is a neutral or charged Higgs boson and V stands for the vector
gauge bosons W or Z, are obtained \cite{Pilaftsis4}. The gauge
invariant forms of the vertices should be $$factor\times [(k\cdot
q)g_{\mu\nu}-k_{\mu}q_{\nu}]$$ where $q_\mu$ and $k_\nu$ are the
four momenta of the photon and vector gauge boson (W or Z) and the
factor depends on the heavy degrees of freedom, the two-loop
Barr-Zee type diagrams only induce the nonzero contributions to
the coefficient $C_5^\pm$.

Setting $I=J=2$ in Eq.\ref{eff1}, we obtain the muon anomalous
magnetic dipole moment in the MSSM:
\begin{eqnarray}
&&\delta a_\mu^{^{SUSY}}={G_Fm_{\mu}^2 \over
\sqrt{2}\pi^2}\Big[ C_2^++C_2^-+{1\over
2}(C_5^-+C_5^+)\Big]_{I=J=2}\;. \label{magnetic}
\end{eqnarray}

Correspondingly, the muon electric dipole moment is
\begin{eqnarray}
&&d_\mu^{^{SUSY}}=e{\sqrt{2}G_Fm_{_\mu}\over i4\pi^2}
\Big[C_5^--C_5^+\Big]_{I=J=2}.
\label{edm}
\end{eqnarray}

For the FCNC process $\tau\rightarrow \mu\gamma$, the
contributions from SM and Higgs sector are suppressed by the small
ratio $x_{_{\nu_i}}=\frac{m_{_{\nu_i}}^2} {m_{_{\rm W}}^2}$. The
supersymmetric contribution originates from the sneutrino-chargino
loop and the amplitude reads:
\begin{eqnarray}
{\cal A}_{_{\tau\rightarrow \mu\gamma}}=
-\frac{eG_F}{4\sqrt{2}\pi^2}\bigg(m_{_{\mu}}F_{_{
\tau\rightarrow\mu\gamma}}^L
\bar{\mu}[/\!\!\!q,/\!\!\!\epsilon]\omega_-
\tau+m_{_{\tau}}F_{_{\tau\rightarrow \mu\gamma}}^R
\bar{\mu}[/\!\!\!q,/\!\!\!\epsilon]\omega_+\tau\bigg)\;,
\label{taumug}
\end{eqnarray}
where $\epsilon$ is the polarization of the emitted photon.
The form factors $F_{_{\tau\rightarrow\mu\gamma}}^L\;,
F_{_{\tau\rightarrow\mu\gamma}}^R$ are formulated as
\begin{eqnarray}
&&F_{_{\tau\rightarrow\mu\gamma}}^L=C_2^-+C_5^-+{m_{_{\tau}}
\over m_{_{\mu}}}C_2^+\;,\nonumber\\
&&F_{_{\tau\rightarrow\mu\gamma}}^R=C_2^++C_5^++{m_{_{\mu}}
\over m_{_{\tau}}}C_2^-\;.
\label{flr}
\end{eqnarray}
From  Eq.\ref{taumug}, we have the decay
width as
\begin{equation}
\Gamma_{_{\tau\rightarrow \mu\gamma}}=\frac{e^2G_F^2m_{_{\tau}}^3}
{128\pi^5}\bigg[m_{_{\mu}}^2\Big|F_{_{\tau\rightarrow
\mu\gamma}}^L\Big|^2+m_{_{\tau}}^2\Big|F_{_{\tau\rightarrow
\mu\gamma}}^R\Big|^2\bigg].
\end{equation}

\section{The lepton-flavor-violating decay $\tau\rightarrow 3\mu$}

\begin{figure}
\setlength{\unitlength}{1mm}
\begin{center}
\begin{picture}(230,200)(55,90)
\put(50,30){\includegraphics{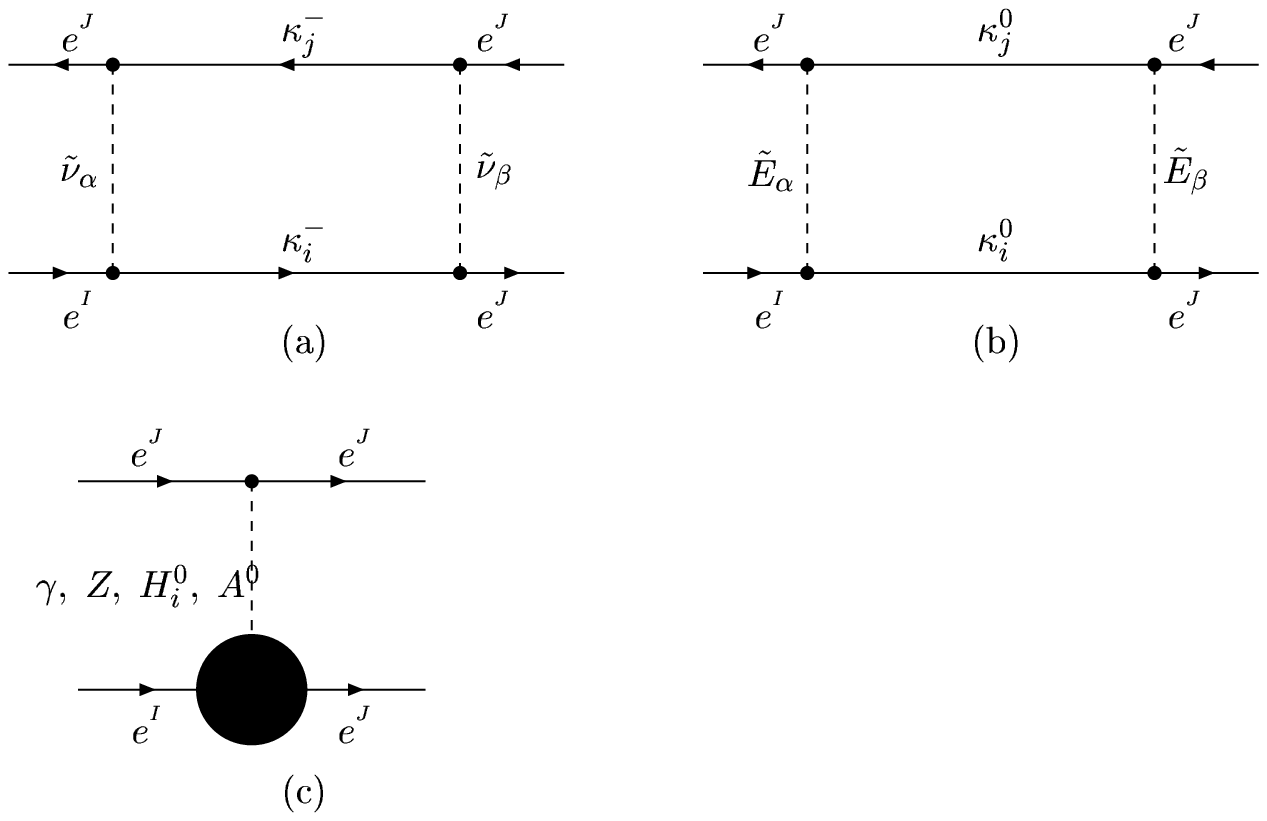}}
\end{picture}
\caption[]{The Feynman diagrams which contribute to the
$e^{^I}\rightarrow 3e^{^J}$ in  MSSM} \label{fig3}
\end{center}
\end{figure}

The effective Lagrangian for $\tau\rightarrow 3\mu$ is induced by
the following four pieces: $\gamma,\;Z,\;H$-mediating penguin and
box diagrams. Those Feynman diagrams are shown in Fig. \ref{fig2}
and Fig. \ref{fig3}. After integrating out these heavy degrees of
freedom, the effective Lagrangian is written as
\begin{eqnarray}
&&{\cal L}_{eff}^{\tau\rightarrow 3\mu}={G_F^2m_{_{\rm W}}^2\over 2\pi^2}
\sum_i C_i {\cal Q}_i
\label{tau3mu}
\end{eqnarray}
with those operators being
\begin{eqnarray}
&&{\cal Q}_1=\bar{\mu}\gamma_\rho\omega_-\tau\bar{\mu}\gamma^\rho\omega_-\mu\;,
\nonumber\\
&&{\cal Q}_2=\bar{\mu}\gamma_\rho\omega_+\tau\bar{\mu}\gamma^\rho\omega_+\mu\;,
\nonumber\\
&&{\cal Q}_3=\bar{\mu}\omega_-\tau\bar{\mu}\omega_-\mu\;,
\nonumber\\
&&{\cal Q}_4=\bar{\mu}\omega_-\tau\bar{\mu}\omega_+\mu\;,
\nonumber\\
&&{\cal Q}_5=\bar{\mu}\omega_+\tau\bar{\mu}\omega_-\mu\;,
\nonumber\\
&&{\cal Q}_6=\bar{\mu}\omega_+\tau\bar{\mu}\omega_+\mu\;.
\nonumber\\
\label{ob2}
\end{eqnarray}
The differential width of $\tau\rightarrow 3\mu$ is
\begin{eqnarray}
&&{d^2\Gamma\over dm_{12}^2dm_{23}^2}={G_F^4m_{_{\rm W}}^4\over (2\pi)^7}
{1\over 32m_{_\tau}^3}|{\cal M}|^2\;,
\label{dwidth}
\end{eqnarray}
where $m_{ij}^2=(p_i+p_j)^2$, and $p_i\;(i=1,\;2,\;3)$ are the
momenta of the outgoing muons in the rest frame of $\tau$. With
Eq. (\ref{tau3mu}), one obtains the square of the transition
matrix element $|{\cal M}|^2$  as
\begin{eqnarray}
&&|{\cal M}|^2=\Big\{4\Big(m_{_\tau}^2+m_{_\mu}^2-m_{12}^2\Big)
\Big(m_{12}^2-2m_{_\mu}^2\Big)\Big(|C_1|^2+|C_2|^2\Big)
\nonumber\\
&&\hspace{1.5cm}+4\Big(m_{_\tau}^2+m_{_\mu}^2-m_{23}^2\Big)
\Big(m_{23}^2-2m_{_\mu}^2\Big)\Big(|C_3|^2+|C_4|^2+|C_5|^2
+|C_6|^2\Big)\nonumber\\
&&\hspace{1.5cm}+8m_{_\tau}m_{_\mu}^3{\bf Re}\Big(4C_1C_2^\dagger+
C_3C_6^\dagger+C_4C_5^\dagger\Big)+4m_{_\mu}^2\Big(m_{_\tau}^2
+m_{_\mu}^2-m_{12}^2\Big){\bf Re}\Big(C_1C_3^\dagger+
C_2C_6^\dagger\Big)\nonumber\\
&&\hspace{1.5cm}+4m_{_\mu}^2\Big(m_{_\tau}^2+m_{_\mu}^2
-m_{13}^2\Big){\bf Re}\Big(C_1C_4^\dagger+C_2C_5^\dagger\Big)
\nonumber\\
&&\hspace{1.5cm}+4m_{_\tau}m_{_\mu}\Big(m_{_\tau}^2+m_{_\mu}^2-m_{12}^2
-m_{23}^2\Big){\bf Re}\Big(C_1C_5^\dagger+C_2C_4^\dagger\Big)
\nonumber\\
&&\hspace{1.5cm}+4m_{_\tau}m_{_\mu}\Big(m_{12}^2-2m_{_\mu}^2\Big)
{\bf Re}\Big(C_1C_6^\dagger+C_2C_4^\dagger\Big)\nonumber\\
&&\hspace{1.5cm}+4m_{_\mu}^2\Big(m_{_\tau}^2+m_{_\mu}^2
-m_{23}^2\Big){\bf Re}\Big(C_3C_4^\dagger+C_5C_6^\dagger\Big)
\nonumber\\
&&\hspace{1.5cm}+4m_{_\tau}m_{_\mu}\Big(m_{23}^2-2m_{_\mu}^2\Big)
{\bf Re}\Big(C_3C_5^\dagger+C_4C_6^\dagger\Big)\Big\}\;.
\label{recmat}
\end{eqnarray}

\section{Numerical result and discussion}

In this section, we present our numerical analysis on the muon
MDM, EDM and the lepton-flavor-violating decay processes in the
supersymmetric scenario with the nonuniversal soft-supersymmetry
breaking. In the lepton sector of MSSM, there are $15+3G$ ($G$
denotes the generation number) new free parameters besides the SM
parameters $g_1,\;g_2,\;m_{_{e^I}}\;\;(I=1,\;2,\;3)$. Too many
parameters reduce the predictability of the model and before any
direct evidence of the SUSY particles is found, determining the
free parameters is the most subtle and tough task in the
supersymmetric theory. To find a way out, the grand unified theory
(GUT) assumption is frequently adopted where all new physics
parameters are fully fixed  from only five free parameters at the
Grand Unification scale.  In that scenario, all the input masses
conserve flavors, namely there are no off-diagonal masses are to
be introduced. Thus when all the concerned parameters evolve from
the GUT scale down to the electroweak scale by the renormalization
group equations (RGE), the flavor changing parts  in the effective
Lagrangian emerge only through the CKM mechanism. The induced
FCNC-related decay modes would have very small branching ratios
which are much lower than the experimental bounds. Therefore, we
would rather adopt an alternative parameterization which may imply
a different physics picture from the GUT scenario, i.e. to choose
a set of parameters at the electroweak scale as inputs. The
numerical results depend on these parameters and by imposing the
experimental bounds, one can obtain constraints on the parameters.
Just as in Ref.\cite{Pilaftsis1}, we require the phase
$arg(\mu)=0$ in order to suppress the one-loop contribution to the
electron EDM.  At the lowest order we can choose the parameter
basis  which is responsible for changing lepton flavors as
following:
$$\mu,\;m_1,\;m_2,\;\tan\beta,\;m_{_{\tilde{E}_I}}^2,\;m_{_{\tilde{E}_{(3+I)}}}^2,\;
\theta_{_{\tilde{E}_I}},\;\phi_{_{\tilde{E}_I}},\;m_{_{L^{IJ}}}^2,\;
m_{_{R^{IJ}}}^2,\;A_{_{IJ}}^l,\;(I,\;J=1,\;2,\;3;\;I\ne J).$$
Correspondingly, the square mass of sneutrino  is given through
the relation:
$$m_{_{\tilde{\nu}_I}}^2=\cos^2\theta_{_{\tilde{E}_I}}m_{_{\tilde{E}_I}}^2
+\sin^2\theta_{_{\tilde{E}_I}}m_{_{\tilde{E}_{(3+I)}}}^2-m_{_{e^I}}^2
+m_{_{\rm W}}^2\cos 2\beta.$$

There are several relevant $CP$ phases as :
$\phi_{\tilde{\mu}}=\phi_{\tilde{L}^I}\;\; {\rm for}\;\; I=2
\;\;\; \phi_{\tilde{\tau}}=\phi_{\tilde{L}^I} \;\;{\rm for}\;\;
I=3,$ and $\phi_{A_{IJ}^l}$, $\phi_{m_{L_{IJ}}^2}$,
$\phi_{m_{R_{IJ}}^2}$ ($I\neq J$) are the $CP$ phases of parameters
$A_{IJ}^l$, $m_{L_{IJ}}^2$ and $m_{R_{IJ}}^2$ respectively.
To simplify our discussion, we assume that the bilinear and
trilinear couplings of scalar quarks are all universal, i.e.
$A_{_{IJ}}^u=A_{_{u^I}}\delta_{IJ},\;A_{_{IJ}}^d=A_{_{d^I}}\delta_{IJ}$
and $m_{_{\tilde{U}_{IJ}}}^2=m_{_{\tilde{U}_{I}}}^2\delta_{IJ},\;
m_{_{\tilde{D}_{IJ}}}^2=m_{_{\tilde{D}_{I}}}^2\delta_{IJ},\;
m_{_{\tilde{Q}_{IJ}}}^2=m_{_{\tilde{Q}_{I}}}^2\delta_{IJ}$. As
aforementioned, the nontrivial $CP$ phases lead to a large mixing
among the neutral Higgs fields\cite{Pilaftsis3}. Considering the
two-loop Yukawa and QCD corrections to the effective potential,
the square mass matrix for the neutral Higgs is written as:
\begin{eqnarray}
&&\hspace{-0.5cm}m_{_{H^0}}^2=\left({\tiny\begin{array}{llc}
\left(\begin{array}{l}
m_a^2s_{_\beta}^2-{8m_{_{\rm W}}^2s_{_{\rm W}}^2\over e^2}
\Big[\lambda_1c_{_\beta}^2\\
+{\bf Re}(\lambda_5)s_{_\beta}^2+
{\bf Re}(\lambda_6)s_{_\beta}c_{_\beta}\Big]
\end{array}\right)
&\left(\begin{array}{l}-m_a^2s_{_\beta}c_{_\beta}
-{8m_{_{\rm W}}^2s_{_{\rm W}}^2\over e^2}\Big[\Big(\lambda_3+\lambda_4
\Big)s_{_\beta}c_{_\beta}\\+{\bf Re}(\lambda_6)c_{_\beta}^2
+{\bf Re}(\lambda_7)s_{_\beta}^2\Big]
\end{array}\right)
&\left(\begin{array}{l}{\bf Im}(\lambda_5)
s_{_\beta}\\+{\bf Im}(\lambda_6)c_{_\beta}\end{array}\right)\\\\
\left(\begin{array}{l}-m_a^2s_{_\beta}c_{_\beta}
-{8m_{_{\rm W}}^2s_{_{\rm W}}^2\over e^2}\Big[\Big(\lambda_3+\lambda_4
\Big)s_{_\beta}c_{_\beta}\\+{\bf Re}(\lambda_6)c_{_\beta}^2
+{\bf Re}(\lambda_7)s_{_\beta}^2\end{array}\right)&
\left(\begin{array}{l}m_a^2c_{_\beta}^2
-{8m_{_{\rm W}}^2s_{_{\rm W}}^2\over e^2}
\Big[\lambda_2s_{_\beta}^2\\+{\bf Re}(\lambda_5)c_{_\beta}^2+
{\bf Re}(\lambda_7)s_{_\beta}c_{_\beta}\Big]\end{array}\right)
&\left(\begin{array}{l}{\bf Im}(\lambda_5)s_{_\beta}\\
+{\bf Im}(\lambda_6)c_{_\beta}\end{array}\right)\\\\
\hspace{0.3cm}{\bf Im}(\lambda_5)s_{_\beta}+{\bf Im}(\lambda_6)c_{_\beta}&
\hspace{0.3cm}{\bf Im}(\lambda_5)c_{_\beta}+{\bf Im}(\lambda_7)s_{_\beta}&
m_a^2
\end{array}}\right)
\label{nhmass}
\end{eqnarray}
with the square mass $m_a^2$
\begin{equation}
m_a^2=m_{_{H^\pm}}^2-{4m_{_{\rm W}}^2s_{_{\rm W}}^2\over e^2}\Big(
{1\over 2}\lambda_4-{\bf Re}(\lambda_5)\Big)\;. \label{ma}
\end{equation}
Here, the parameter $m_{_{H^\pm}}$ represents the  mass of the
physical charged Higgs-bosons, and the concrete expressions of the
other parameters $\lambda_i \;(i=1,2,\cdots,7)$ are presented in
appendix \ref{apd2}. As pointed out by the authors of Ref.
\cite{Pilaftsis3}, the experimental lower bound on the mass of the
lightest neutral Higgs mass can be reduced to 60 ${\rm GeV}$ in
the CP-violating MSSM. In the numerical analysis of this work, we
take this lower bound as an input.
With above specification about the parameter space, we carry out
our numerical computations. Without losing generality, we take
$m_{_{H^\pm}} =300\;{\rm GeV},\;m_1=1\;{\rm
TeV},\;m_{_{\tilde{e}_{1}}}=m_{_{\tilde{e}_{2}}}=10\;{\rm
TeV},\;m_{_{\tilde{\mu}_{1}}}=m_{_{\tilde{\tau}_{1}}}=500\;{\rm
GeV},\;m_{_{\tilde{\mu}_{2}}}=m_{_{\tilde{\tau}_{2}}}=700\;{\rm
GeV},\;\theta_{_{\tilde e}}=\phi_{_{\tilde
e}}=0,\;\theta_{_{\tilde \mu}} =\theta_{_{\tilde \tau}}={\pi\over
4},\;m_{_{L^{12}}}^2=
m_{_{L^{13}}}^2=m_{_{R^{12}}}^2=m_{_{R^{13}}}^2=0\;{\rm GeV
}^2,\;A_{12}^l=A_{13}^l=A_{21}^l=A_{31}^l=0\;{\rm GeV},\; M_{_{\rm
SUSY}}=1\;{\rm TeV},\;m_t=174\;{\rm GeV},\;m_b=4.5\;{\rm GeV}
,\;m_{_{\tilde{Q}^3}}=m_{_{\tilde{U}^3}}=m_{_{\tilde{D}^3}}=
500\;{\rm GeV},\;A_t=A_b=e^{i{\pi\over 4}}\;{\rm TeV}$ all through
the paper. Indeed, the electron EDM  is a very rigorous constraint
to the parameters. Although we set the CP phase which is only
related to the first generation of sleptons to be zero (in this
case the one-loop contribution to the electron EDM  is also zero),
the non-zero CP phases which are related to the second and third
generations can also lead to a large EDM of electron via the
two-loop Braa-Zee diagrams. For some regions in the parameter
space, the two-loop Barr-Zee diagrams would contribute an EDM of
electron which is larger than the experimental upper bound. Thus
the  electron EDM  restricts the relevant parameters via the
two-loop diagrams. Considering the constraint of the electron  EDM
through two-loop Barr-Zee diagrams involving sleptons, we set the
$CP$ phases $\phi_{_{\tilde \mu}} =\phi_{_{\tilde \tau}}={\pi\over
18}$.

\begin{figure}
\setlength{\unitlength}{1mm}
\begin{center}
\begin{picture}(230,200)(55,90)
\put(50,30){\includegraphics{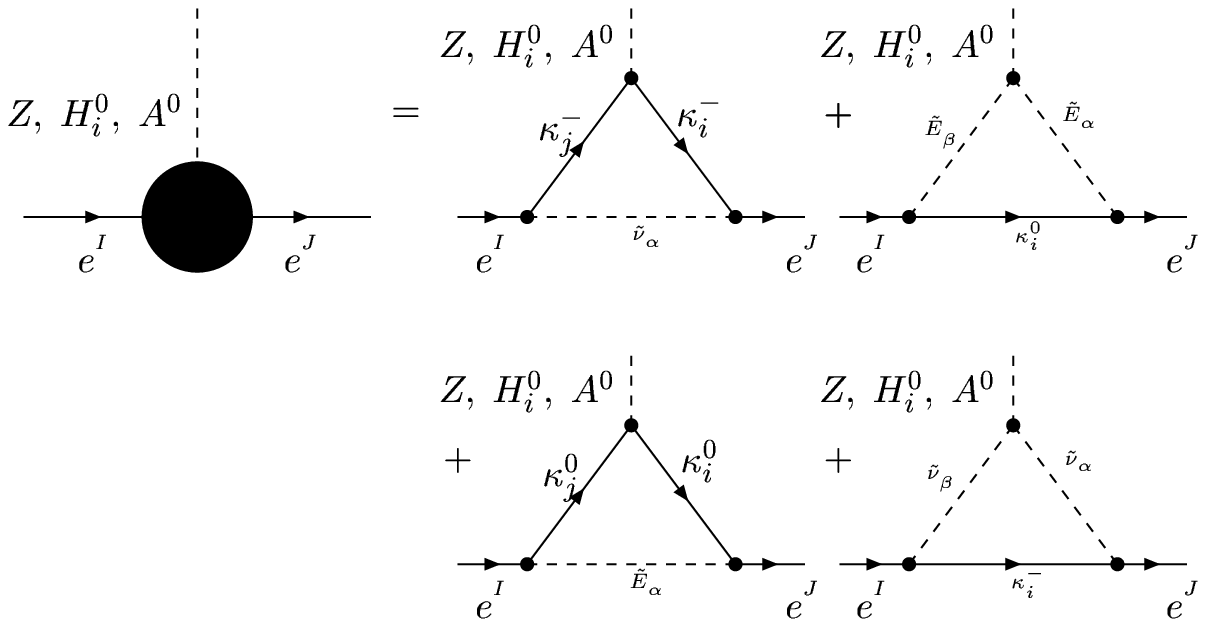}}
\end{picture}
\caption[]{The penguin diagrams  in  MSSM where $Z\;,H_i^0,\;A^0$
are involved respectively } \label{fig4}
\end{center}
\end{figure}

\begin{figure}
\setlength{\unitlength}{1mm}
\begin{center}
\begin{picture}(230,200)(55,90)
\put(50,80){\includegraphics{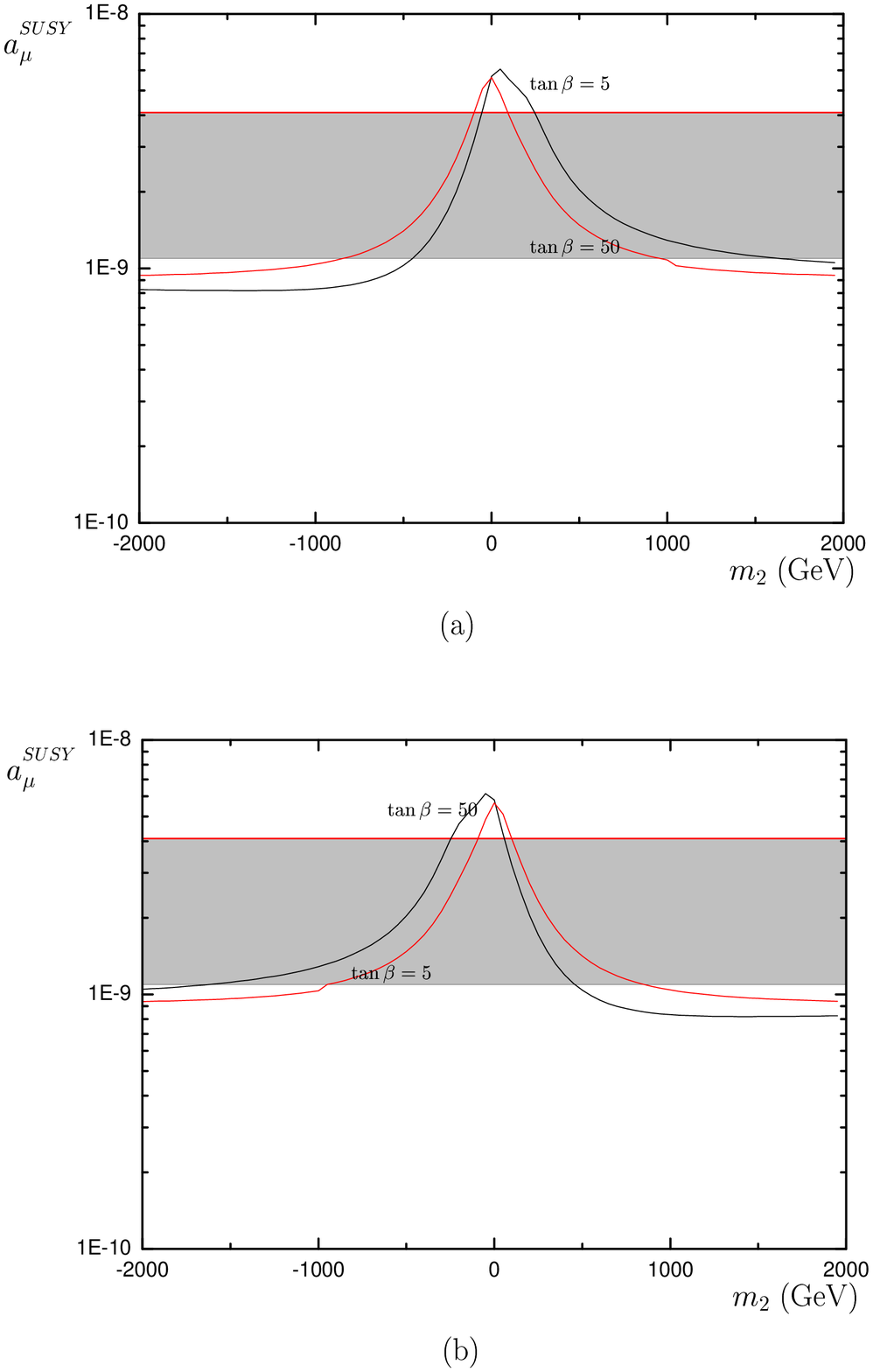}}
\end{picture}
\caption[]{The muon anomalous magnetic dipole moment versus $m_2$
in  MSSM with (a)$\mu=-200\;{\rm GeV}$, (b)$\mu=200\;{\rm GeV}$
and $\tan\beta=5,\;50$, the other parameters are set as in the
text. The shaded region is allowed by $1\sigma$ tolerance from the
most recent experimental observation.} \label{fig5}
\end{center}
\end{figure}

At present, the experimental upper bounds on the branching ratios
of the two lepton-flavor-violating processes are
$$B(\tau\rightarrow\mu\gamma)<1.1\times 10^{-6},\;\; B(\tau\rightarrow
3\mu)<1.9\times 10^{-6}.$$ Upon imposition of the experimental
limits on our theoretical predictions for the branching ratios and
the muon MDM, we find that the contribution to the muon MDM from
slepton generation mixing parameters is much less than the
contribution from the flavor conserving parameters through
scanning the parameter space. Taking
$\phi_{_{A^l_{23}}}=\phi_{_{A^l_{32}}}=
\phi_{_{m^2_{L_{23}}}}=\phi_{_{m^2_{R_{23}}}}={\pi\over 2}$, and
$|m^2_{L_{23}}|=|m^2_{R_{23}}|=100\;{\rm
GeV}^2,\;|A^l_{23}|=|A^l_{23}|=100\;{\rm GeV}$, we plot the muon
MDM versus the parameter $m_2$ in Fig. \ref{fig4}. Within
$1\sigma$ tolerance, we find that the theoretical prediction
coincides with the experimental data rather well if a suitable
parameter range is adopted. As for the calculation on the muon
EDM, we must consider the constraint from the  electron EDM. With
$2\sigma$ tolerance, the experimental upper bound on the electron
EDM is $|d_e|<0.5\times 10^{-26}\;{e\cdot cm}$. With
$m_2=500\;{\rm GeV},\;\phi_{_{A^l_{23}}}=\phi_{_{A^l_{32}}}=
\phi_{_{m^2_{L_{23}}}}=\phi_{_{m^2_{R_{23}}}}={\pi\over 2}$, and
$|m^2_{L_{23}}|=|m^2_{R_{23}}|=100\;{\rm
GeV}^2,\;|A^l_{23}|=|A^l_{23}|=100\;{\rm GeV}$, and
$\tan\beta=5,\,50$ respectively, we plot the electron EDM, the
muon EDM, and the muon MDM  versus the parameter $\mu$ in Fig.
\ref{fig5}. We find that the muon EDM can reach $10^{-23}\;(e\cdot
cm)$, which is one order above the proposed sensitivity of the
coming experiments \cite{Semertzidis}: $10^{-24}\;(e\cdot cm),$
while the muon MDM and the electron EDM are also consistent with
the experimental data.

Now, we analyze the lepton-flavor-violating decays of heavy-lepton
$\tau$: $\tau\rightarrow\mu\gamma$, $\tau\rightarrow 3\mu$. With
$\mu=200\;{\rm GeV}$ and set all CP phases to be zero, we plot the
branching ratio of $\tau\rightarrow 3\mu$ versus
$m_{_{L_{23}}}^2=m_{_{R_{23}}}^2=\delta m^2$ (Fig. \ref{fig7}(a))
and $A^l_{23}=A^l_{23}=\delta A$ (Fig. \ref{fig7}(b)). This result
is slightly different from that given in previous literature. We
not only consider the contributions from the $\gamma-$penguin,
$Z-$penguin and the box diagrams, but also include the
Higgs-penguin. We notice that for larger $\tan\beta$ values, the
contribution of the Higgs-penguin is not negligible, It is in
analog to its role for the rare leptonic decays of B-meson.
Numerically, we find that the contribution of the $\gamma-$penguin
is much less than that of the $Z-$penguin and box diagrams. For
the case of non-zero $A_{23}^l$,$A_{32}^l$ and large $\tan\beta$,
the Higgs-penguin is also non-negligible. When
$A_{23}^l=A_{32}^l=0$, but $m_{_{L_{23}}}^2$, and
$m_{_{R_{23}}}^2$ are not zero, the main contribution to the decay
width of $\tau\rightarrow 3\mu$ comes from the the $Z-$penguin and
box diagrams. By this we can explain why the difference of the
curves corresponding to $\tan\beta=5$ and 50, is rather small,
whereas in Fig.7(b) the difference is so obvious. As to
$\tau\rightarrow\mu\gamma$, when we take into account the
constraint from $\tau\rightarrow 3\mu,$ we find that its branching
ratio is obviously lower than the experimental bound. If the
lepton flavor violation originates from the term
$m_{_{L_{ij}}}^2,\;m_{_{R_{ij}}}^2$ (Fig. \ref{fig7}(a)), the
difference between $\tan\beta=5$ and $\tan\beta=50$ is very small.
If the lepton flavor violation originates from the term $A^l_{ij}$
(Fig. \ref{fig7}(b)), the difference between $\tan\beta=5$ and
$\tan\beta=50$ is about two orders. We also investigate the
$\tau\rightarrow\mu\gamma$ in the model. Imposing the constraint
on the parameters from $\tau\rightarrow 3\mu$, we find that the
branching ratio for $\tau\rightarrow\mu\gamma$ is smaller than
$10^{-8}$, beyond the present experimental detecting ability.
Assuming that the flavor violation originates from
$m_{_{L^{ij}}}^2,\;m_{_{R^{ij}}}^2$, we plot the
$B(\tau\rightarrow\mu\gamma)$ within the possible parameter range
in Fig. \ref{fig8}. The situation is  similar to the case of
$\tau\rightarrow 3\mu$, namely if the flavor violation originates
from the term $A^l_{ij}$, the branching ratio
$B(\tau\rightarrow\mu\gamma)$ is much less than that if the mixing
$m_{_{L^{ij}}}^2,\;m_{_{R^{ij}}}^2$ induces the flavor violation.

\section{Conclusion}

In this work, we investigate the muon anomalous magnetic dipole
moment, muon electric dipole moment, and the branching ratios of
$\tau\rightarrow\mu\gamma$ and $\tau\rightarrow 3\mu$ in the
framework of the CP-violating MSSM with  the nonuniversal
soft-supersymmetry breaking. For the muon anomalous magnetic
dipole moment and electric dipole moment, the main contribution
comes from the parameters which conserve the flavors. Process
$\tau\rightarrow 3\mu$ occurs mainly through the
$Z-,\;H_i^0-,\;A^0-$ penguins and box diagrams. It can help
understanding why the branching ratio of
$\tau\rightarrow\mu\gamma$ is so small as long as we consider the
constraints from $\tau\rightarrow 3\mu$. From the methodology
aspect, our method is equivalent to the MIA scheme when all mass
eigenvalues of the mass matrix are almost degenerate, or the
square mass difference among the eigenvalues are much larger than
the corresponding flavor violation parameters. For the case where
only several mass eigenvalues are degenerate, the MIA is invalid.
Indeed, our method is an improvement of the MIA.

\vspace{1.0cm} \noindent {\Large{\bf Acknowledgments}}

This work is partially supported by the National Natural Science
Foundation of China. One of the authors (T.-F. Feng) is also
partly supported by the K. C. Wong Post-doctoral Research
Fund, Hongkong.

\appendix

\section{The mixing matrix for the sfermion sector \label{app1}}

For the general case, the mixing matrices for the sfermions are
given in the text. In this appendix, we present the mixing matrix
for the case where there is degeneracy in the mass spectra. In fact,
we can always perform a transformation on the sfermion mass matrix and turn it into
a standard form:
\begin{eqnarray}
&&{\cal Z}_{_{S_{LR}}}m_{_{S}}^2{\cal Z}_{_{S_{LR}}}^{-1}=\left(
\begin{array}{cccccc}
m_{_{S_1}}^2&\Delta m_{_{S_{12}}}^2&\Delta
m_{_{S_{13}}}^2&0&\Delta m_{_{S_{15}}}^2&\Delta
m_{_{S_{16}}}^2\\
\Delta m_{_{S_{12}}}^{2*}&m_{_{S_2}}^2&\Delta
m_{_{S_{23}}}^2&\Delta m_{_{S_{24}}}^2&0&\Delta
m_{_{S_{26}}}^2\\
\Delta m_{_{S_{13}}}^{2*}&\Delta
m_{_{S_{23}}}^{2*}&m_{_{S_3}}^2&\Delta m_{_{S_{34}}}^2&\Delta
m_{_{S_{35}}}^2&0\\
0&\Delta m_{_{S_{24}}}^{2*}&\Delta
m_{_{S_{34}}}^{2*}&m_{_{S_4}}^2&\Delta m_{_{S_{45}}}^2&\Delta
m_{_{S_{46}}}^2\\
\Delta m_{_{S_{15}}}^{2*}&0&\Delta m_{_{S_{35}}}^{2*}&\Delta
m_{_{S_{45}}}^{2*}&m_{_{S_5}}^2&\Delta m_{_{S_{56}}}^2\\
\Delta m_{_{S_{16}}}^{2*}&\Delta m_{_{S_{26}}}^{2*}&0&\Delta
m_{_{S_{46}}}^{2*}&\Delta m_{_{S_{56}}}^{2*}&m_{_{S_6}}^2
\end{array}\right)\;,
\label{mixdeg}
\end{eqnarray}
with $\Delta m_{_{S_{ij}}}^2$ being given in Eq. (\ref{slodm}) and
the matrix ${\cal Z}_{_{S_{LR}}}$ is formulated as
\begin{eqnarray}
&&{\cal Z}_{_{S_{LR}}}=\left(\begin{array}{cccccc}
\cos\theta_{_{S_1}}&0&0&\sin\theta_{_{S_1}}e^{-i\phi_{_{S_1}}}&0&0\\
0&\cos\theta_{_{S_2}}&0&0&\sin\theta_{_{S_2}}e^{-i\phi_{_{S_2}}}&0\\
0&0&\cos\theta_{_{S_3}}&0&0&\sin\theta_{_{S_3}}e^{-i\phi_{_{S_3}}}\\
-\sin\theta_{_{S_1}}e^{i\phi_{_{S_1}}}&0&0&\cos\theta_{_{S_1}}&0&0\\
0&-\sin\theta_{_{S_2}}e^{i\phi_{_{S_2}}}&0&0&\cos\theta_{_{S_2}}&0\\
0&0&-\sin\theta_{_{S_3}}e^{i\phi_{_{S_2}}}&0&0&\cos\theta_{_{S_3}}
\end{array}\right)\;.\nonumber\\
\label{mm1}
\end{eqnarray}
Without losing generality, we assume that the first $n\;(2\le n\le
5)$ eigenvalues satisfy the condition $m^2_{_{{\tilde{S}_{i}}}}
-m^2_{_{{\tilde{S}_{j}}}}\sim \Delta
m^2_{_{{\tilde{S}_{ij}}}}\;\;(i,\;j=1,\;\cdots,\;n)$, for the
scalar lepton sector the mixing matrix is given as
\begin{eqnarray}
&&{\cal Z}_{_{S}}={\cal Z}_{_{S_{LR}}}{\cal Z}_{_{\bf A}}{\bf V}
\label{mm2}
\end{eqnarray}
with
\begin{eqnarray}
&&{\bf V}_{_{{S}_{ii}}}=1-\sum\limits_{j\neq i}\sum\limits_{\alpha
}{|\Delta x_{_{{S}_{j\alpha}}}\Delta x_{_{{S}_{\alpha
i}}}|^2\over(x_{_{{S}_i}}-x_{_{{S}_j}})^2
(x_{_{{S}_i}}-x_{_{{S}_\alpha}})^2}-\sum\limits_{\alpha }{|\Delta
x_{_{{S}_{i\alpha}}}|^2\over(x_{_{{S}_i}}
-x_{_{{S}_\alpha}})^2}\;,
\nonumber\\
&&{\bf V}_{_{{S}_{ji}}}=\sum\limits_{\alpha} {\Delta
x_{_{{S}_{j\alpha}}}\Delta x_{_{{S}_{\alpha i}}}\over
(x_{_{{S}_i}}-x_{_{{S}_j}})(x_{_{{S}_i}}
-x_{_{{S}_\alpha}})}-{1\over (x_{_{{S}_i}}
-x_{_{{S}_j}})^2}\sum\limits_{\alpha,\beta}{|\Delta x_{_{{S}
_{\alpha i}}}|^2\Delta x_{_{{S}_{j\beta}}}\Delta x_{_{{S} _{\beta
i}}}\over (x_{_{{S}_i}}-x_{_{{S}_\alpha}})
(x_{_{{S}_\beta}}-x_{_{{S}_i}})}\;,\nonumber\\
&&{\bf V}_{_{{S}_{\alpha i}}}={\Delta x_{_{{S}_{\alpha i}}} \over
x_{_{{S}_i}}-x_{_{{S}_\alpha}}} +\sum\limits_{\beta\neq
\alpha}{\Delta x_{_{{S}_{\alpha\beta}}} \Delta x_{_{{S}_{\beta
i}}}\over (x_{_{{S}_i}}
-x_{_{{S}_\alpha}})(x_{_{{S}_i}}-x_{_{{S}_\beta}})}
+\sum\limits_{j}\sum\limits_{\beta}{\Delta x_{_{{S}_{\alpha j}}}
\Delta x_{_{{S}_{j\beta}}}\Delta x_{_{{S}_{\beta i}}} \over
(x_{_{{S}_i}}-x_{_{{S}_j}})( x_{_{{S}_i}}-x_{_{{S}_\alpha}})(
x_{_{{S}_i}}-x_{_{{S}_\beta}})}\;,\nonumber\\
&&{\bf V}_{_{{S}_{i\alpha}}}={\Delta x_{_{{S}_{i\alpha}}} \over
x_{_{{S}_\alpha}}-x_{_{{S}_i}}}
+\sum\limits_{\beta\neq\alpha}{\Delta x_{_{{S}_{i\beta}}} \Delta
x_{_{{S}_{\beta\alpha}}}\over (x_{_{{S}_\alpha}}
-x_{_{{S}_i}})(x_{_{{S}_\alpha}}-x_{_{{S}_\beta}})}
\;,\nonumber\\
&&{\bf V}_{_{{S}_{\alpha\alpha}}}=1-\sum\limits_{\beta }{|\Delta
x_{_{{S}_{\alpha\beta}}}|^2\over(x_{_{{S}_\alpha}}
-x_{_{{S}_\beta}})^2}\;,
\nonumber\\
&&{\bf V}_{_{{S}_{\beta\alpha}}}={\Delta x_{_{{S}_{\beta\alpha}}}
\over(x_{_{{S}_\alpha}}-x_{_{{S}_\beta}})} +\sum\limits_{i}{\Delta
x_{_{{S}_{\beta i}}} \Delta
x_{_{{S}_{i\alpha}}}\over(x_{_{{S}_\alpha}}
-x_{_{{S}_\beta}})(x_{_{{S}_\beta}}-x_{_{{S}_i}})}
+\sum\limits_{\gamma\neq\alpha,\beta}{\Delta
x_{_{{S}_{\beta\gamma}}} \Delta x_{_{{S}_{\gamma\alpha}}}\over
(x_{_{{S}_\alpha}}
-x_{_{{S}_\beta}})(x_{_{{S}_\alpha}}-x_{_{{S}_\gamma}})} \;.
\label{mm3}
\end{eqnarray}
and
\begin{eqnarray}
{\cal Z}_{_{\bf A}}=\left(\begin{array}{ll}{\bf A}_{_{n\times
n}}&{\bf 0}_{_{n\times(6-n)}}\\{\bf 0}_{_{(6-n)\times n}}&{\bf
1}_{_{(6-n)\times(6-n)}}
\end{array}\right). \label{mm4}
\end{eqnarray}
Here, the matrix ${\bf A}$ is used to diagonalize the block
matrix $m_{_{S_{ij}}}^2\;\;(i,\;j=1,\cdots,n)$. Expanding the Eq.
(\ref{mm2}), we have
\begin{eqnarray}
&&{\cal Z}_{_{S}}^{\dagger \alpha i}={\bf U}_{_{\tilde{E}_{\alpha
i}}}\cos\theta_{_{\tilde{E}_i}}+{\bf
U}_{_{\tilde{E}_{\alpha(3+i)}}}\sin\theta_{_{\tilde{E}_i}}
e^{-i\varphi_{_{\tilde{E}_i}}}\;,
\nonumber\\
&&{\cal Z}_{_{S}}^{\dagger\alpha(3+i)}= -{\bf
U}_{_{\tilde{E}_{\alpha i}}}\sin\theta_{_{\tilde{E}_i}}
e^{i\varphi_{_{\tilde{E}_i}}} +{\bf
U}_{_{\tilde{E}_{\alpha(3+i)}}}\cos\theta_{_{\tilde{E}_i}}
\;\;(\alpha=1,\cdots,6;\;i=1,2,3)\;, \label{mm5}
\end{eqnarray}
where
\begin{eqnarray}
&&{\bf U}_{_{{S}_{ij}}}=\sum\limits_{k}{\bf V}_{_{{S}_{ik}}}
{\bf A}_{_{{S}_{kj}}}\;,\nonumber\\
&&{\bf U}_{_{{S}_{i\alpha}}}={\bf V}_{_{{S}_{i\alpha}}}
\;,\nonumber\\
&&{\bf U}_{_{{S}_{\alpha i}}}=\sum\limits_{k} {\bf
V}_{_{{S}_{\alpha k}}}{\bf A}_{_{{S}_{ki}}}\;,
\nonumber\\
&&{\bf U}_{_{{S}_{\alpha\beta}}}={\bf V}_{_{{S}_{\alpha\beta}}}
\;.
\end{eqnarray}

\section{Expressions of the coefficients in Eq.(\ref{eff1})
for the lepton anomalous magnetic dipole moment and the amplitude
of decay mode $e^I\rightarrow e^J\gamma$ \label{app2}}

The expressions of the coefficients in Eq.(\ref{eff1}) for the vertex
$e^Ie^J\gamma$ are
\begin{eqnarray}
&&C_1^-(\mu_{_{\rm W}})=-2e
{\bf G}_{_{\nu^{(a)}}}^{^{\{iiI\alpha J\alpha\}}}
x_{_{\kappa_i^-}}^2{\cal B}_{_{[41]}}^{^{(0)}}(x_{_{\kappa_i^-}},
x_{_{\tilde{\nu}^\alpha}})\nonumber\\&&\hspace{2.0cm}
+{1\over c_{_{\rm W}}^2}
{\bf G}_{_{L^{(a)}}}^{^{\{iiI\alpha J\alpha\}}}
{\cal F}_1(x_{_{\tilde{E}^\alpha}},x_{_{\kappa_i^0}})
\;,\nonumber\\
&&C_2^-(\mu_{_{\rm W}})=-{1\over 2}
{\bf G}_{_{\nu^{(a)}}}^{^{\{iiI\alpha J\alpha\}}}
{\cal F}_1(x_{_{\kappa_i^-}},x_{_{\tilde{\nu}^\alpha}})
\nonumber\\&&\hspace{2.0cm}
-{1\over 4c_{_{\rm W}}^2}
{\bf G}_{_{L^{(a)}}}^{^{\{iiI\alpha J\alpha\}}}
{\cal F}_1(x_{_{\tilde{E}^\alpha}},x_{_{\kappa_i^0}})
\;,\nonumber\\
&&C_3^-(\mu_{_{\rm W}})=-{2\over 3}
{\bf G}_{_{\nu^{(a)}}}^{^{\{iiI\alpha J\alpha\}}}
\Big[{\cal B}_{_{[21]}}^{^{(0)}}(x_{_{\kappa_i^-}},
x_{_{\tilde{\nu}^\alpha}})
-{5\over 2}x_{_{\kappa_i^-}}{\cal B}_{_{[31]}}^{^{(0)}}(
x_{_{\kappa_i^-}},x_{_{\tilde{\nu}^\alpha}})
\nonumber\\&&\hspace{2.0cm}
-{7\over 2}x_{_{\kappa_i^-}}^2{\cal B}_{_{[41]}}^{^{(0)}}(
x_{_{\kappa_i^-}},x_{_{\tilde{\nu}^\alpha}})
\Big]-{1\over 6c_{_{\rm W}}^2}
{\bf G}_{_{L^{(a)}}}^{^{\{iiI\alpha J\alpha\}}}
{\cal B}_{_{[41]}}^{^{(2)}}(x_{_{\tilde{E}^\alpha}},x_{_{\kappa_i^0}})
\;,\nonumber\\
&&C_4^-(\mu_{_{\rm W}})=-{\sqrt{2}\over c_\beta}
{\bf G}_{_{\nu^{(b)}}}^{^{\{iiI\alpha J\alpha\}}}x_{_{\kappa_i^-}}^{3\over 2}
x_{_{e^J}}^{1\over 2}{\cal B}_{_{[31]}}^{^{(0)}}(
x_{_{\kappa_i^-}},x_{_{\tilde{\nu}^\alpha}})
\nonumber\\&&\hspace{2.0cm}
-{1\over c_{_{\rm W}}^2}
{\bf G}_{_{L^{(b)}}}^{^{\{iiI\alpha J\alpha\}}}
x_{_{\tilde{E}^\alpha}}x_{_{\kappa_i^0}}^{1\over 2}
{\cal B}_{_{[31]}}^{^{(0)}}(x_{_{\tilde{E}^\alpha}},x_{_{\kappa_i^0}})
\;,\nonumber\\
&&C_5^-(\mu_{_{\rm W}})=-{1\over \sqrt{2}c_\beta}
{\bf G}_{_{\nu^{(b)}}}^{^{\{iiI\alpha J\alpha\}}}
(x_{_{\kappa_i^-}}x_{_{e^J}})^{1\over 2}
{\cal B}_{_{[31]}}^{^{(1)}}(x_{_{\kappa_i^-}},x_{_{\tilde{\nu}^\alpha}})
\nonumber\\&&\hspace{2.0cm}
+{1\over 2c_{_{\rm W}}^2}
{\bf G}_{_{L^{(b)}}}^{^{\{iiI\alpha J\alpha\}}}
x_{_{\tilde{E}^\alpha}}x_{_{\kappa_i^0}}^{1\over 2}
{\cal B}_{_{[31]}}^{^{(0)}}(x_{_{\tilde{E}^\alpha}},x_{_{\kappa_i^0}})
\;,\nonumber\\
&&C_1^+(\mu_{_{\rm W}})=-{1\over c_\beta^2}
{\bf G}_{_{\nu^{(d)}}}^{^{\{iiI\alpha J\alpha\}}}
(x_{_{e^I}}x_{_{e^J}})^{1\over 2}x_{_{\kappa_i^-}}^2
{\cal B}_{_{[41]}}^{^{(0)}}(x_{_{\kappa_i^-}},x_{_{\tilde{\nu}^\alpha}})
\nonumber\\&&\hspace{2.0cm}
+{1\over c_{_{\rm W}}^2}
{\bf G}_{_{L^{(d)}}}^{^{\{iiI\alpha J\alpha\}}}
{\cal F}_1(x_{_{\tilde{E}^\alpha}},x_{_{\kappa_i^0}})
\;,\nonumber\\
&&C_2^+(\mu_{_{\rm W}})=-{1\over 4c_\beta^2}
{\bf G}_{_{\nu^{(d)}}}^{^{\{iiI\alpha J\alpha\}}}
(x_{{e^I}}x_{_{e^J}})^{1\over 2}
{\cal F}_1(x_{_{\kappa_i^-}},x_{_{\tilde{\nu}^\alpha}})
\nonumber\\&&\hspace{2.0cm}
-{1\over 4c_{_{\rm W}}^2}
{\bf G}_{_{L^{(d)}}}^{^{\{iiI\alpha J\alpha\}}}
{\cal F}_1(x_{_{\tilde{E}^\alpha}},x_{_{\kappa_i^0}})
\;,\nonumber\\
&&C_3^+(\mu_{_{\rm W}})=-{1\over 3c_\beta^2}
{\bf G}_{_{\nu^{(d)}}}^{^{\{iiI\alpha J\alpha\}}}
(x_{_{e^I}}x_{_{e^J}})^{1\over 2}\Big[
{\cal B}_{_{[21]}}^{^{(0)}}(x_{_{\kappa_i^-}},x_{_{\tilde{\nu}^\alpha}})
-{5\over 2}x_{_{\kappa_i^-}}{\cal B}_{_{[31]}}^{^{(0)}}(
x_{_{\kappa_i^-}},x_{_{\tilde{\nu}^\alpha}})
\nonumber\\&&\hspace{2.0cm}
-{7\over 2}x_{_{\kappa_i^-}}^2{\cal B}_{_{[41]}}^{^{(0)}}(
x_{_{\kappa_i^-}},x_{_{\tilde{\nu}^\alpha}})
\Big]-{1\over 6c_{_{\rm W}}^2}
{\bf G}_{_{L^{(d)}}}^{^{\{iiI\alpha J\alpha\}}}
{\cal B}_{_{[41]}}^{^{(2)}}(x_{_{\tilde{E}^\alpha}},x_{_{\kappa_i^0}})
\;,\nonumber\\
&&C_4^+(\mu_{_{\rm W}})=-{\sqrt{2}\over c_\beta}
{\bf G}_{_{\nu^{(c)}}}^{^{\{iiI\alpha J\alpha\}}}
x_{_{\kappa_i^-}}^{3\over 2}x_{_{e^I}}^{1\over 2}
{\cal B}_{_{[31]}}^{^{(0)}}(x_{_{\kappa_i^-}},x_{_{\tilde{\nu}^\alpha}})
\nonumber\\&&\hspace{2.0cm}
-{1\over c_{_{\rm W}}^2}
{\bf G}_{_{L^{(c)}}}^{^{\{iiI\alpha J\alpha\}}}
x_{_{\tilde{E}^\alpha}}x_{_{\kappa_i^0}}^{1\over 2}
{\cal B}_{_{[31]}}^{^{(0)}}(x_{_{\tilde{E}^\alpha}},x_{_{\kappa_i^0}})
\;,\nonumber\\
&&C_5^+(\mu_{_{\rm W}})=-{1\over \sqrt{2}
c_\beta}{\bf G}_{_{\nu^{(c)}}}^{^{\{iiI\alpha J\alpha\}}}
(x_{_{\kappa_i^-}}x_{_{e^I}})^{1\over 2}
{\cal B}_{_{[31]}}^{^{(1)}}(x_{_{\kappa_i^-}},x_{_{\tilde{\nu}^\alpha}})
\nonumber\\&&\hspace{2.0cm}
+{1\over 2c_{_{\rm W}}^2}
{\bf G}_{_{L^{(c)}}}^{^{\{iiI\alpha J\alpha\}}}
x_{_{\tilde{E}^\alpha}}x_{_{\kappa_i^0}}^{1\over 2}
{\cal B}_{_{[31]}}^{^{(0)}}(x_{_{\tilde{E}^\alpha}},x_{_{\kappa_i^0}})
\;.\label{eegam}
\end{eqnarray}

\section{The couplings among the sfermions and Higgs\label{apd1}}

In this appendix, we give the couplings among the scalar fermions
and Higgs fields that involve the two-loop Barr-Zee-type contributions
to the Wilson coefficients:
\begin{eqnarray}
&&\Gamma^{H^+\tilde{U}_i\tilde{D}_j}=
\Big\{{e\over \sqrt{2}s_{_{\rm W}}}V^{IJ*}
\Big[\Big(-2m_{_{\rm W}}s_{_\beta}c_{_\beta}+{m_{_{d^J}}^2\over m_{_{\rm W}}}
\tan\beta+{m_{_{u^I}}^2\over m_{_{\rm W}}\tan\beta}\Big)
{\cal Z}_{_{\tilde U}}^{Ii}{\cal Z}_{_{\tilde D}}^{Jj}
\nonumber\\&&\hspace{2.0cm}+{m_{_{u^I}}m_{_{d^J}}\over m_{_{\rm W}}
s_{_\beta}c_{_\beta}}{\cal Z}_{_{\tilde U}}^{(3+I)i}
{\cal Z}_{_{\tilde D}}^{(3+J)j}\Big]
+\Big({e\mu^*m_{_{u^I}}\over \sqrt{2}
m_{_{\rm W}}s_{_{\rm W}}}V^{IJ*}-V^{KJ*}A_{_{KI}}^uc_{_\beta}\Big)
{\cal Z}_{_{\tilde U}}^{(3+I)i}{\cal Z}_{_{\tilde D}}^{Jj}
\nonumber\\&&\hspace{2.0cm}+\Big({e\mu m_{_{d^J}}\over \sqrt{2}
m_{_{\rm W}}s_{_{\rm W}}}V^{IJ*}+V^{IK*}A_{_{KJ}}^ds_{_\beta}\Big)
{\cal Z}_{_{\tilde U}}^{(3+I)i}{\cal Z}_{_{\tilde D}}^{Jj}
\Big\}\;,\nonumber\\
&&\Gamma^{H^+\tilde{\nu}_i\tilde{E}_j}={\cal Z}_{_\nu}^{Ii}\Big\{
{e\over \sqrt{2}s_{_{\rm W}}}\Big({\mu m_{_{e^I}}\over m_{_{\rm W}}}
{\cal Z}_{_E}^{(3+I)j}-2m_{_{\rm W}}s_{_\beta}c_{_\beta}
{\cal Z}_{_E}^{Ij}\Big)+\Big({em_{_{e^I}}^2\over \sqrt{2}
m_{_{\rm W}}s_{_{\rm W}}}\tan\beta{\cal Z}_{_E}^{Ij}+
s_{_\beta}A_{_{IK}}^l{\cal Z}_{_E}^{(3+K)j}\Big)\Big\}\;,\nonumber\\
&&\Gamma^{A\tilde{U}_i\tilde{U}_j}=i\Big\{{em_{_{u^I}}\over 2
m_{_{\rm W}}s_{_{\rm W}}}\Big(\mu{\cal Z}_{_U}^{Ii*}{\cal Z}_{_U}^{(3+I)j}
-\mu^*{\cal Z}_{_U}^{Ij}{\cal Z}_{_U}^{(3+I)i*}\Big)
+{c_{_\beta}\over\sqrt{2}}\Big(A_{_{IJ}}^u
{\cal Z}_{_U}^{Ij}{\cal Z}_{_U}^{(3+J)i*}-A_{_{IJ}}^{u*}
{\cal Z}_{_U}^{Ii*}{\cal Z}_{_U}^{(3+J)j}\Big)\Big\}\;,\nonumber\\
&&\Gamma^{A\tilde{D}_i\tilde{D}_j}=i\Big\{
{s_{_\beta}\over\sqrt{2}}\Big(A_{_{IJ}}^{d*}
{\cal Z}_{_D}^{Ij}{\cal Z}_{_D}^{(3+J)i*}-A_{_{IJ}}^{d}
{\cal Z}_{_D}^{Ii*}{\cal Z}_{_D}^{(3+J)j}\Big)-{em_{_{d^I}}\over 2
m_{_{\rm W}}s_{_{\rm W}}}\Big(\mu{\cal Z}_{_D}^{Ii*}{\cal Z}_{_D}^{(3+I)j}
-\mu^*{\cal Z}_{_D}^{Ij}{\cal Z}_{_D}^{(3+I)i*}\Big)\Big\}\;,\nonumber\\
&&\Gamma^{A\tilde{E}_i\tilde{E}_j}=i\Big\{
{s_{_\beta}\over\sqrt{2}}\Big(A_{_{IJ}}^{l*}
{\cal Z}_{_E}^{Ij}{\cal Z}_{_E}^{(3+J)i*}-A_{_{IJ}}^{l}
{\cal Z}_{_E}^{Ii*}{\cal Z}_{_E}^{(3+J)j}\Big)-{em_{_{e^I}}\over 2
m_{_{\rm W}}s_{_{\rm W}}}\Big(\mu{\cal Z}_{_E}^{Ii*}{\cal Z}_{_E}^{(3+I)j}
-\mu^*{\cal Z}_{_E}^{Ij}{\cal Z}_{_E}^{(3+I)i*}\Big)\Big\}\;.
\label{sfhiggs}
\end{eqnarray}
\section{Expressions of the coefficients in Eq.(\ref{tau3mu}) for
decay mode $e^{^I}\rightarrow 3e^{^J}$\label{app3}}

The contributions from the gamma penguin diagrams are written as
\begin{eqnarray}
&&C_1^{\gamma}(\mu_{_{\rm W}})={4s_{_{\rm W}}^2\over 3}
{\bf G}_{_{\nu^{(a)}}}^{^{\{iiI\alpha J\alpha\}}}
\Big[{\cal B}_{_{[21]}}^{^{(0)}}(x_{_{\kappa_i^-}},
x_{_{\tilde{\nu}^\alpha}})
-{5\over 2}x_{_{\kappa_i^-}}{\cal B}_{_{[31]}}^{^{(0)}}(
x_{_{\kappa_i^-}},x_{_{\tilde{\nu}^\alpha}})
\nonumber\\&&\hspace{2.0cm}
-{7\over 2}x_{_{\kappa_i^-}}^2{\cal B}_{_{[41]}}^{^{(0)}}(
x_{_{\kappa_i^-}},x_{_{\tilde{\nu}^\alpha}})
\Big]+{s_{_{\rm W}}^2\over 3c_{_{\rm W}}^2}
{\bf G}_{_{L^{(a)}}}^{^{\{iiI\alpha J\alpha\}}}
{\cal B}_{_{[41]}}^{^{(2)}}(x_{_{\tilde{E}^\alpha}},x_{_{\kappa_i^0}})
\;,\nonumber\\
&&C_2^{\gamma}(\mu_{_{\rm W}})={2s_{_{\rm W}}^2\over 3
c_\beta^2}{\bf G}_{_{\nu^{(d)}}}^{^{\{iiI\alpha J\alpha\}}}
(x_{_{e^I}}x_{_{e^J}})^{1\over 2}\Big[
{\cal B}_{_{[21]}}^{^{(0)}}(x_{_{\kappa_i^-}},x_{_{\tilde{\nu}^\alpha}})
-{5\over 2}x_{_{\kappa_i^-}}{\cal B}_{_{[31]}}^{^{(0)}}(
x_{_{\kappa_i^-}},x_{_{\tilde{\nu}^\alpha}})
\nonumber\\&&\hspace{2.0cm}
-{7\over 2}x_{_{\kappa_i^-}}^2{\cal B}_{_{[41]}}^{^{(0)}}(
x_{_{\kappa_i^-}},x_{_{\tilde{\nu}^\alpha}})
\Big]+{s_{_{\rm W}}^2\over 3c_{_{\rm W}}^2}
{\bf G}_{_{L^{(d)}}}^{^{\{iiI\alpha J\alpha\}}}
{\cal B}_{_{[41]}}^{^{(2)}}(x_{_{\tilde{E}^\alpha}},x_{_{\kappa_i^0}})
\;,\nonumber\\
&&C_4^{\gamma}(\mu_{_{\rm W}})=-2C_1^{\gamma}(\mu_{_{\rm W}})\;,\nonumber\\
&&C_5^{\gamma}(\mu_{_{\rm W}})=-2C_2^{\gamma}(\mu_{_{\rm W}})\;,\nonumber\\
&&C_3^{\gamma}(\mu_{_{\rm W}})=C_6^{\gamma}(\mu_{_{\rm W}})=0\;.\nonumber\\
\label{gp}
\end{eqnarray}
The pieces from the $Z-$penguin diagrams are
\begin{eqnarray}
&&C_1^Z(\mu_{_{\rm W}})={\cos^2 2\theta_{_{\rm W}}\over
c_{_{\rm W}}^2}{\bf G}_{_{L^{(a)}}}
^{^{\{iiI\alpha J\alpha\}}}{\cal F}_2(x_{_{\kappa_i^0}},
x_{_{\tilde{E}^\alpha}})\nonumber\\&&\hspace{2.0cm}
+{\cos 2\theta_{_{\rm W}}\over 4c_{_{\rm W}}^2}
{\bf G}_{_{L^{(a)}}}^{^{\{ijI\alpha J\alpha\}}}\Big({\cal Z}_{_N}^{4i*}
{\cal Z}_{_N}^{4j}-{\cal Z}_{_N}^{3i}{\cal Z}_{_N}^{3j*}\Big)\Big[{1\over 2}
{\cal T}_{_{[111]}}^{^{(1)}}(x_{_{\tilde{E}^\alpha}},x_{_{\kappa_i^0}},
x_{_{\kappa_j^0}})\nonumber\\
&&\hspace{2.0cm}+(x_{_{\kappa_i^0}}x_{_{\kappa_j^0}})^{1\over 2}
{\cal T}_{_{[111]}}^{^{(0)}}(x_{_{\tilde{E}^\alpha}},x_{_{\kappa_i^0}},
x_{_{\kappa_j^0}})\Big]\nonumber\\
&&\hspace{2.0cm}+{\cos 2\theta_{_{\rm W}}\over 2c_{_{\rm W}}^2}
{\bf G}_{_{L^{(a)}}}^{^{\{iiI\beta J\alpha\}}}\Big(2s_{_{\rm W}}^2
\delta^{\alpha\beta}-{\cal Z}_{_{\tilde{E}}}^{K\alpha}
{\cal Z}_{_{\tilde{E}}}^{K\beta*}\Big){\cal T}_{_{[111]}}^{^{(1)}}(x_{_{\tilde{E}^\alpha}},
x_{_{\tilde{E}^\beta}},x_{_{\kappa_i^0}})
\nonumber\\&&\hspace{2.0cm}
+{2\cos^2 2\theta_{_{\rm W}}\over c_{_{\rm W}}^2}
{\bf G}_{_{\nu^{(a)}}}^{^{\{iiI\alpha J\alpha\}}}
{\cal F}_2(X_{_{\kappa_i^0}},x_{_{\tilde{\nu}^\alpha}})
\nonumber\\
&&\hspace{2.0cm}+\cos 2\theta_{_{\rm W}}
{\bf G}_{_{\nu^{(a)}}}^{^{\{ijI\alpha J\alpha\}}}
\Big[{1\over 2}{\cal T}_{_{[111]}}^{^{(1)}}(x_{_{\tilde{\nu}^\alpha}},x_{_{\kappa_i^-}}
,x_{_{\kappa_j^-}})\Big(2\delta^{ij}
\cos 2\theta_{_{\rm W}}+{\cal Z}_-^{1i}{\cal Z}_-^{1j*}\Big)
\nonumber\\
&&\hspace{2.0cm}-(x_{_{\kappa_i^-}}x_{_{\kappa_j^-}})^{1\over 2}
{\cal T}_{_{[111]}}^{^{(0)}}(x_{_{\tilde{\nu}^\alpha}},x_{_{\kappa_i^-}},
x_{_{\kappa_j^-}})\Big(2\delta^{ij}\cos
2\theta_{_{\rm W}}+{\cal Z}_+^{1i*}{\cal Z}_+^{1j}\Big)\Big]
\nonumber\\&&\hspace{2.0cm}
+\cos 2\theta_{_{\rm W}}{\bf G}_{_{L^{(a)}}}^{^{\{iiI\beta J\alpha\}}}
{\cal T}_{_{[111]}}^{^{(1)}}(x_{_{\tilde{\nu}^\alpha}},
x_{_{\tilde{\nu}^\beta}},x_{_{\kappa_i^-}})\;,
\nonumber\\
&&C_2^Z(\mu_{_{\rm W}})={4s_{_{\rm W}}^4\over c_{_{\rm W}}^2}
{\bf G}_{_{L^{(d)}}}^{^{\{iiI\alpha J\alpha\}}}
{\cal F}_2(x_{_{\kappa_i^0}},x_{_{\tilde{E}^\alpha}})
\nonumber\\
&&\hspace{2.0cm}+{s_{_{\rm W}}^2\over
2c_{_{\rm W}}^2}\Big({\cal Z}_{_N}^{4i*}
{\cal Z}_{_N}^{4j}-{\cal Z}_{_N}^{3i}{\cal Z}_{_N}^{3j*}\Big)
{\bf G}_{_{L^{(d)}}}^{^{\{ijI\alpha J\alpha\}}}
\Big[{1\over 2}{\cal T}_{_{[111]}}^{^{(1)}}(x_{_{\tilde{E}^\alpha}},
x_{_{\kappa_i^0}},x_{_{\kappa_j^0}})\nonumber\\
&&\hspace{2.0cm}-(x_{_{\kappa_i^0}}x_{_{\kappa_j^0}})^{1\over 2}
{\cal T}_{_{[111]}}^{^{(0)}}(x_{_{\tilde{E}^\alpha}},x_{_{\kappa_i^0}},
x_{_{\kappa_j^0}})\Big]\nonumber\\
&&\hspace{2.0cm}-{s_{_{\rm W}}^2\over c_{_{\rm W}}^2}\Big(2s_{_{\rm W}}^2
\delta^{\alpha\beta}-{\cal Z}_{_{\tilde{E}}}^{K\alpha}
{\cal Z}_{_{\tilde{E}}}^{K\beta*}\Big){\bf G}_{_{L^{(d)}}}^{^{\{iiI\beta J\alpha\}}}
{\cal T}_{_{[111]}}^{^{(1)}}(
x_{_{\tilde{E}^\alpha}},x_{_{\tilde{E}^\beta}},x_{_{\kappa_i^0}})\nonumber\\
&&\hspace{2.0cm}+{4s_{_{\rm W}}^4\over
c_\beta^2}(x_{_{e^I}}x_{_{e^J}})^{1\over 2}
{\bf G}_{_{\nu^{(d)}}}^{^{\{iiI\alpha J\alpha\}}}
{\cal F}_2(x_{_{\kappa_i^-}},x_{_{\tilde{\nu}^\alpha}})\nonumber\\
&&\hspace{2.0cm}-{s_{_{\rm W}}^2
\over c_\beta^2}(x_{_{e^I}}x_{_{e^J}})^{1\over 2}
{\bf G}_{_{\nu^{(d)}}}^{^{\{ijI\alpha J\alpha\}}}\Big[{1\over 2}
{\cal T}_{_{[111]}}^{^{(1)}}(x_{_{\tilde{\nu}^\alpha}},
x_{_{\kappa_i^-}},x_{_{\kappa_j^-}})\Big(2\delta^{ij}
\cos 2\theta_{_{\rm W}}+{\cal Z}_+^{1i*}{\cal Z}_+^{1j}\Big)
\nonumber\\
&&\hspace{2.0cm}-(x_{_{\kappa_i^-}}x_{_{\kappa_j^-}})^{1\over 2}
{\cal T}_{_{[111]}}^{^{(0)}}(x_{_{\tilde{\nu}^\alpha}},
x_{_{\kappa_i^-}},x_{_{\kappa_j^-}})\Big(2\delta^{ij}\cos
2\theta_{_{\rm W}}+{\cal Z}_-^{1i}{\cal Z}_-^{1j*}\Big)\Big]
\nonumber\\&&\hspace{2.0cm}
-{2s_{_{\rm W}}^2\over c_\beta^2}{\bf G}_{_{\nu^{(d)}}}^{^{\{iiI\beta J\alpha\}}}
(x_{_{e^I}}x_{_{e^J}})^{1\over 2}{\cal T}_{_{[111]}}^{^{(1)}}(
x_{_{\tilde{\nu}^\alpha}},x_{_{\tilde{\nu}^\beta}},x_{_{\kappa_i^-}})
\;,\nonumber\\
&&C_4^Z(\mu_{_{\rm W}})={4s_{_{\rm W}}^2\over \cos 2\theta_{_{\rm W}}}
C_1^Z(\mu_{_{\rm W}})\;,\nonumber\\
&&C_5^Z(\mu_{_{\rm W}})={\cos 2\theta_{_{\rm W}}\over s_{_{\rm W}}^2}
C_2^Z(\mu_{_{\rm W}})\;,\nonumber\\
&&C_3^Z(\mu_{_{\rm W}})=C_6^Z(\mu_{_{\rm W}})=0\;.
\label{zp}
\end{eqnarray}
The $CP$-even Higgs penguin-diagram contributions are formulated as
\begin{eqnarray}
&&C_1^{H^0_k}(\mu_{_{\rm W}})=C_2^{H^0_k}(\mu_{_{\rm W}})=0\;,\nonumber\\
&&C_3^{H^0_k}(\mu_{_{\rm W}})={1\over 2c_{_{\rm W}}^2c_\beta^2x_{_{H_k^0}}}
{\cal F}_2(x_{_{\kappa_i^0}},x_{_{\tilde{E}^\alpha}})
\Big({\cal Z}_{_R}^{1k}\Big)^2
\Big[(x_{_{e^I}}x_{_{e^J}})^{1\over 2}{\bf G}_{_{L^{(d)}}}^{^{\{iiI\alpha J\alpha\}}}
+x_{_{e^J}}{\bf G}_{_{L^{(a)}}}^{^{\{iiI\alpha J\alpha\}}}\Big]\nonumber\\
&&\hspace{2.0cm}+{1\over 2c_{_{\rm W}}^3c_\beta
x_{_{H_k^0}}}x_{_{e^I}}^{1\over 2}{\cal Z}_{_R}^{1k}
{\bf G}_{_{L^{(b)}}}^{^{\{ijI\alpha J\alpha\}}}
\Big[{\cal T}_{_{[111]}}^{^{(1)}}(x_{_{\tilde{E}^\alpha}},x_{_{\kappa_i^0}}
,x_{_{\kappa_j^0}})\Big({\cal Z}_{_R}^{1k}{\cal Z}_{_N}^{3i*}
-{\cal Z}_{_R}^{2k}{\cal Z}_{_N}^{4i*}\Big)
\nonumber\\&&\hspace{2.0cm}
\Big({\cal Z}_{_N}^{1j*}s_{_{\rm W}}-{\cal Z}_{_N}^{2j*}c_{_{\rm W}}\Big)
+(x_{_{\kappa_i^0}}x_{_{\kappa_j^0}})^{1\over 2}
{\cal T}_{_{[111]}}^{^{(0)}}(x_{_{\tilde{E}^\alpha}},x_{_{\kappa_i^0}}
,x_{_{\kappa_j^0}})\Big({\cal Z}_{_R}^{1k}{\cal Z}_{_N}^{3j}\nonumber\\
&&\hspace{2.0cm}-{\cal Z}_{_R}^{2k}{\cal Z}_{_N}^{4j}\Big)
\Big({\cal Z}_{_N}^{1i}s_{_{\rm W}}-{\cal Z}_{_N}^{2i}c_{_{\rm W}}\Big)
\Big]\nonumber\\&&\hspace{2.0cm}
+{s_{_{\rm W}}\over em_{_{\rm W}}c_{_{\rm W}}^2c_\beta x_{_{H_k^0}}
}(x_{_{e^I}}x_{_{\kappa_i^0}})^{1\over 2}{\cal Z}_{_R}^{1k}
{\bf E}{\bf G}_{_{L^{(b)}}}^{^{\{iiI\beta J\alpha\}}}
{\cal T}_{_{[111]}}^{^{(0)}}(x_{_{\kappa_i^0}},x_{_{\tilde{E}^\alpha}},
x_{_{\tilde{E}^\beta}})\nonumber\\
&&\hspace{2.0cm}
+{1\over c_\beta^2x_{_{H_k^0}}}
{\cal F}_2(x_{_{\kappa_i^-}},x_{_{\tilde{\nu}^\alpha}})\Big({\cal Z}_{_R}^{1k}\Big)^2
\Big[(x_{_{e^I}}x_{_{e^J}})^{1\over 2}
{\bf G}_{_{\nu^{(a)}}}^{^{\{iiI\alpha J\alpha\}}}\nonumber\\
&&\hspace{2.0cm}+{(x_{_{e^I}}^3x_{_{e^J}})^{1\over 2}\over 2
c_\beta^2}{\bf G}_{_{\nu^{(d)}}}^{^{\{iiI\alpha J\alpha\}}}\Big]
\nonumber\\&&\hspace{2.0cm}
-{2\over c_\beta^2x_{_{H_k^0}}}(x_{_{e^I}}x_{_{e^J}})^{1\over 2}
{\bf G}_{_{\nu^{(b)}}}^{^{\{jiI\alpha J\alpha\}}}
\Big[{\cal T}_{_{[111]}}^{^{(1)}}(
x_{_{\tilde{\nu}^\alpha}},x_{_{\kappa_i^-}},x_{_{\kappa_j^-}})
\Big({\cal Z}_{_R}^{1k}{\cal Z}_-^{2j*}{\cal Z}_+^{1i*}
\nonumber\\&&\hspace{2.0cm}+
{\cal Z}_{_R}^{2k}{\cal Z}_-^{1j*}{\cal Z}_+^{2i*}\Big)+
(x_{_{\kappa_i^-}}x_{_{\kappa_j^-}})^{1\over 2}{\cal T}_{_{[111]}}^{^{(0)}}(
x_{_{\tilde{\nu}^\alpha}},x_{_{\kappa_i^-}},x_{_{\kappa_j^-}})
\Big({\cal Z}_{_R}^{1k}{\cal Z}_-^{2i}{\cal Z}_+^{1j}
\nonumber\\&&\hspace{2.0cm}+
{\cal Z}_{_R}^{2k}{\cal Z}_-^{1i}{\cal Z}_+^{2j}\Big)\Big]
{\cal Z}_{_R}^{1k}
\nonumber\\&&\hspace{2.0cm}
-{e\over 2\sqrt{2}m_{_{\rm W}}
s_{_{\rm W}}c_{_{\rm W}}^2c_\beta^2x_{_{H_k^0}}}(x_{_{e^I}}x_{_{e^J}}
x_{_{\kappa_i^-}})^{1\over 2}{\cal Z}_{_R}^{1k}B_{_R}^k
{\bf G}_{_{\nu^{(b)}}}^{^{\{iiI\alpha J\alpha\}}}
{\cal B}_{_{[21]}}^{^{(0)}}(x_{_{\tilde{\nu}^\alpha}},x_{_{\kappa_i^-}})
\;,\nonumber\\
&&C_4^{H^0_k}(\mu_{_{\rm W}})=C_3^{H^0_k}(\mu_{_{\rm W}})\;,\nonumber\\
&&C_5^{H^0_k}(\mu_{_{\rm W}})={1\over 2c_{_{\rm W}}^2c_\beta^2x_{_{H_k^0}}}
{\cal F}_2(x_{_{\kappa_i^0}},x_{_{\tilde{E}^\alpha}})
\Big({\cal Z}_{_R}^{1k}\Big)^2
\Big[x_{_{e^J}}{\bf G}_{_{\nu^{(d)}}}^{^{\{iiI\alpha J\alpha\}}}
+(x_{_{e^I}}x_{_{e^J}})^{1\over 2}
{\bf G}_{_{\nu^{(a)}}}^{^{\{iiI\alpha J\alpha\}}}
\Big]\nonumber\\
&&\hspace{2.0cm}+{2\over 2c_{_{\rm W}}^3c_\beta x_{_{H_k^0}}}x_{_{e^I}}^{1\over 2}
{\cal Z}_{_R}^{1k}{\bf G}_{_{\nu^{(c)}}}^{^{\{ijI\alpha J\alpha\}}}
\Big[{\cal T}_{_{[111]}}^{^{(1)}}(x_{_{\tilde{E}^\alpha}},
x_{_{\kappa_i^0}},x_{_{\kappa_j^0}})\Big({\cal Z}_{_R}^{1k}{\cal Z}_{_N}^{3j}
-{\cal Z}_{_R}^{2k}{\cal Z}_{_N}^{4j}\Big)
\nonumber\\&&\hspace{2.0cm}
\Big({\cal Z}_{_N}^{1i}s_{_{\rm W}}-{\cal Z}_{_N}^{2i}c_{_{\rm W}}\Big)
+(x_{_{\kappa_i^0}}x_{_{\kappa_j^0}})^{1\over 2}{\cal T}_{_{[111]}}^{^{(0)}}(
x_{_{\tilde{E}^\alpha}},x_{_{\kappa_i^0}},x_{_{\kappa_j^0}})
\Big({\cal Z}_{_R}^{1k}{\cal Z}_{_N}^{3i*}\nonumber\\
&&\hspace{2.0cm}-{\cal Z}_{_R}^{2k}{\cal Z}_{_N}^{4i*}\Big)
\Big({\cal Z}_{_N}^{1j*}s_{_{\rm W}}-{\cal Z}_{_N}^{2j*}c_{_{\rm W}}\Big)
\Big]\nonumber\\&&\hspace{2.0cm}
+{s_{_{\rm W}}\over em_{_{\rm W}}c_{_{\rm W}}^2
c_\beta x_{_{H_k^0}}}(x_{_{e^I}}x_{_{\kappa_i^0}})^{1\over 2}
{\cal Z}_{_R}^{1k}{\bf E}{\bf G}_{_{\nu^{(c)}}}^{^{\{iiI\beta J\alpha\}}}
{\cal T}_{_{[111]}}^{^{(0)}}(x_{_{\kappa_i^0}},
x_{_{\tilde{E}^\alpha}},x_{_{\tilde{E}^\beta}})
\nonumber\\
&&\hspace{2.0cm}
+{1\over c_\beta^2x_{_{H_k^0}}}
{\cal F}_2(x_{_{\kappa_i^-}},x_{_{\tilde{\nu}^\alpha}})
\Big({\cal Z}_{_R}^{1k}\Big)^2
\Big[x_{_{e^I}}{\bf G}_{_{\nu^{(a)}}}^{^{\{iiI\alpha J\alpha\}}}
+{x_{_{e^I}}x_{_{e^J}}\over 2c_\beta^2}
{\bf G}_{_{\nu^{(d)}}}^{^{\{iiI\alpha J\alpha\}}}\Big]
\nonumber\\&&\hspace{2.0cm}
-{2\over c_\beta^2x_{_{H_k^0}}}
x_{_{e^I}}\Big[{\cal T}_{_{[111]}}^{^{(1)}}(x_{_{\tilde{\nu}^\alpha}},
x_{_{\kappa_i^-}},x_{_{\kappa_j^-}})\Big({\cal Z}_{_R}^{1k}{\cal Z}_-^{2i}
{\cal Z}_+^{1j}\nonumber\\&&\hspace{2.0cm}+
{\cal Z}_{_R}^{2k}{\cal Z}_-^{1i}{\cal Z}_+^{2j}\Big)+
(x_{_{\kappa_i^-}}x_{_{\kappa_j^-}})^{1\over 2}
{\cal T}_{_{[111]}}^{^{(0)}}(x_{_{\tilde{\nu}^\alpha}},x_{_{\kappa_i^-}},
x_{_{\kappa_j^-}})\Big({\cal Z}_{_R}^{1k}{\cal Z}_-^{2j*}{\cal Z}_+^{1i*}
\nonumber\\&&\hspace{2.0cm}+
{\cal Z}_{_R}^{2k}{\cal Z}_-^{1j*}{\cal Z}_+^{2i*}\Big)\Big]
{\cal Z}_{_R}^{1k}{\bf G}_{_{\nu^{(c)}}}^{^{\{ijI\alpha J\alpha\}}}
\nonumber\\&&\hspace{2.0cm}
-{e\over 2\sqrt{2}m_{_{\rm W}}
s_{_{\rm W}}c_{_{\rm W}}^2c_\beta^2x_{_{H_k^0}}}x_{_{e^I}}
x_{_{\kappa_i^-}}^{1\over 2}{\bf G}_{_{\nu^{(c)}}}^{^{\{iiI\alpha J\alpha\}}}
{\cal Z}_{_R}^{1k}B_{_R}^k
{\cal B}_{_{[21]}}^{^{(0)}}(x_{_{\tilde{\nu}^\alpha}},x_{_{\kappa_i^-}})
\;,\nonumber\\
&&C_6^{H^0_k}(\mu_{_{\rm W}})=C_5^{H^0_k}(\mu_{_{\rm W}})\;.\nonumber\\
\label{cpehp}
\end{eqnarray}
The $CP$-odd Higgs contributions are
\begin{eqnarray}
&&C_1^{A^0}(\mu_{_{\rm W}})=C_2^{A^0}(\mu_{_{\rm W}})=0\;,\nonumber\\
&&C_3^{A^0}(\mu_{_{\rm W}})={1\over 2c_{_{\rm W}}^2x_{_{A^0}}}\tan^2\beta
{\cal F}_2(x_{_{\kappa_i^0}},x_{_{\tilde{E}^\alpha}})
\Big[(x_{_{e^I}}x_{_{e^J}})^{1\over 2}
{\bf G}_{_{L^{(d)}}}^{^{\{iiI\alpha J\alpha\}}}
+x_{_{e^J}}{\bf G}_{_{L^{(a)}}}^{^{\{iiI\alpha J\alpha\}}}
\Big]\nonumber\\
&&\hspace{2.0cm}+{1\over 2c_{_{\rm W}}^3x_{_{A^0}}}x_{_{e^I}}^{1\over 2}
\tan\beta\Big[-{\cal T}_{_{[111]}}^{^{(1)}}(x_{_{\tilde{E}^\alpha}},
x_{_{\kappa_i^0}},x_{_{\kappa_j^0}})\Big(s_\beta{\cal Z}_{_N}^{3i*}
-c_\beta{\cal Z}_{_N}^{4i*}\Big)\nonumber\\&&\hspace{2.0cm}
\Big({\cal Z}_{_N}^{1j*}s_{_{\rm W}}-{\cal Z}_{_N}^{2j*}c_{_{\rm W}}\Big)
+(x_{_{\kappa_i^0}}x_{_{\kappa_j^0}})^{1\over 2}
{\cal T}_{_{[111]}}^{^{(0)}}(x_{_{\tilde{E}^\alpha}},x_{_{\kappa_i^0}},
x_{_{\kappa_j^0}})\Big(s_\beta{\cal Z}_{_N}^{3j}\nonumber\\
&&\hspace{2.0cm}-c_\beta{\cal Z}_{_N}^{4j}\Big)
\Big({\cal Z}_{_N}^{1i}s_{_{\rm W}}-{\cal Z}_{_N}^{2i}c_{_{\rm W}}\Big)
\Big]{\bf G}_{_{L^{(b)}}}^{^{\{ijI\alpha J\alpha\}}}
\nonumber\\&&\hspace{2.0cm}
+{s_{_{\rm W}}\over em_{_{\rm W}}c_{_{\rm W}}^2
x_{_{A^0}}}(x_{_{e^I}}x_{_{\kappa_i^0}})^{1\over 2}\tan\beta
{\bf F}{\bf G}_{_{L^{(b)}}}^{^{\{iiI\beta J\alpha\}}}
{\cal T}_{_{[111]}}^{^{(0)}}(x_{_{\kappa_i^0}},x_{_{\tilde{E}^\alpha}}
,x_{_{\tilde{E}^\beta}})
\nonumber\\
&&\hspace{2.0cm} +{1\over x_{_{A^0}}} \tan^2\beta{\cal
F}_2(x_{_{\kappa_i^-}},x_{_{\tilde{\nu}^\alpha}})
\Big[x_{_{e^J}}{\bf G}_{_{\nu^{(a)}}}^{^{\{iiI\alpha J\alpha\}}}
+{x_{_{e^I}}x_{_{e^J}}\over 2c_\beta^2}
{\bf G}_{_{\nu^{(d)}}}^{^{\{iiI\alpha J\alpha\}}}\Big]
\nonumber\\&&\hspace{2.0cm}
-{2\over s_\beta x_{_{A^0}}}(x_{_{e^I}}x_{_{e^J}})^{1\over 2}\tan^2\beta
\Big[{\cal T}_{_{[111]}}^{^{(1)}}(x_{_{\tilde{\nu}^\alpha}},
x_{_{\kappa_i^-}},x_{_{\kappa_j^-}})\Big(s_\beta{\cal Z}_-^{2j*}{\cal Z}_+^{1i*}
\nonumber\\&&\hspace{2.0cm}+
c_\beta{\cal Z}_-^{1j*}{\cal Z}_+^{2i*}\Big)-
(x_{_{\kappa_i^-}}x_{_{\kappa_j^-}})^{1\over 2}
{\cal T}_{_{[111]}}^{^{(0)}}(x_{_{\tilde{\nu}^\alpha}},x_{_{\kappa_i^-}}
x_{_{\kappa_j^-}})\Big(s_\beta{\cal Z}_-^{2i}{\cal Z}_+^{1j}
+c_\beta{\cal Z}_-^{1i}{\cal Z}_+^{2j}\Big)\Big]
{\bf G}_{_{L^{(b)}}}^{^{\{jiI\alpha J\alpha\}}}
\;,\nonumber\\
&&C_4^{A^0}(\mu_{_{\rm W}})=-C_3^{A^0}(\mu_{_{\rm W}})\;,\nonumber\\
&&C_5^{A^0}(\mu_{_{\rm W}})=-{1\over 2c_{_{\rm W}}^2x_{_{A^0}}}\tan^2\beta
{\cal F}_2(x_{_{\kappa_i^0}},x_{_{\tilde{E}^\alpha}})
\Big[x_{_{e^J}}{\bf G}_{_{L^{(d)}}}^{^{\{iiI\alpha J\alpha\}}}
+(x_{_{e^I}}x_{_{e^J}})^{1\over 2}
{\bf G}_{_{L^{(a)}}}^{^{\{iiI\alpha J\alpha\}}}\Big]\nonumber\\
&&\hspace{2.0cm}-{1\over 2c_{_{\rm W}}^3x_{_{A^0}}}x_{_{e^I}}^{1\over 2}
\tan\beta{\bf G}_{_{L^{(c)}}}^{^{\{ijI\alpha J\alpha\}}}
\Big[{\cal T}_{_{[111]}}^{^{(1)}}(x_{_{\tilde{E}^\alpha}},
x_{_{\kappa_i^0}},x_{_{\kappa_j^0}})\Big(s_\beta{\cal Z}_{_N}^{3j}
-c_\beta{\cal Z}_{_N}^{4j}\Big)
\nonumber\\&&\hspace{2.0cm}
\Big({\cal Z}_{_N}^{1i}s_{_{\rm W}}-{\cal Z}_{_N}^{2i}c_{_{\rm W}}\Big)
-(x_{_{\kappa_i^0}}x_{_{\kappa_j^0}})^{1\over 2}
{\cal T}_{_{[111]}}^{^{(0)}}(x_{_{\tilde{E}^\alpha}},x_{_{\kappa_i^0}},
x_{_{\kappa_j^0}})\Big(s_\beta{\cal Z}_{_N}^{3i*}\nonumber\\
&&\hspace{2.0cm}-c_\beta{\cal Z}_{_N}^{4i*}\Big)
\Big({\cal Z}_{_N}^{1j*}s_{_{\rm W}}-{\cal Z}_{_N}^{2j*}c_{_{\rm W}}\Big)
\Big]\nonumber\\&&\hspace{2.0cm}
-{s_{_{\rm W}}\over em_{_{\rm W}}
c_{_{\rm W}}^2x_{_{A^0}}}(x_{_{e^I}}x_{_{\kappa_i^0}})^{1\over 2}
\tan\beta{\bf F}{\bf G}_{_{L^{(c)}}}^{^{\{iiI\beta J\alpha\}}}
{\cal T}_{_{[111]}}^{^{(0)}}(x_{_{\kappa_i^0}},
x_{_{\tilde{E}^\alpha}},x_{_{\tilde{E}^\beta}})
\nonumber\\
&&\hspace{2.0cm}
-{1\over x_{_{A^0}}}
\tan^2\beta{\cal F}_2(x_{_{\kappa_i^-}},x_{_{\tilde{\nu}^\alpha}})
\Big[(x_{_{e^I}}x_{_{e^J}})^{1\over 2}
{\bf G}_{_{\nu^{(a)}}}^{^{\{iiI\alpha J\alpha\}}}
+{(x_{_{e^I}}x_{_{e^J}}^3)^{1\over 2} \over 2c_\beta^2}
{\bf G}_{_{\nu^{(d)}}}^{^{\{iiI\alpha J\alpha\}}}\Big]
\nonumber\\&&\hspace{2.0cm}
+{1\over s_\beta x_{_{A^0}}}
x_{_{e^I}}\tan^2\beta{\bf G}_{_{\nu^{(c)}}}^{^{\{ijI\alpha J\alpha\}}}
\Big[{\cal T}_{_{[111]}}^{^{(1)}}(
x_{_{\tilde{\nu}^\alpha}},x_{_{\kappa_i^-}},x_{_{\kappa_j^-}})
\Big(s_\beta{\cal Z}_-^{2i}{\cal Z}_+^{1j}+
c_\beta{\cal Z}_-^{1i}{\cal Z}_+^{2j}\Big)\nonumber\\&&\hspace{2.0cm}-
(x_{_{\kappa_i^-}}x_{_{\kappa_j^-}})^{1\over 2}
{\cal T}_{_{[111]}}^{^{(0)}}(x_{_{\tilde{\nu}^\alpha}},
x_{_{\kappa_i^-}},x_{_{\kappa_j^-}})\Big(s_\beta{\cal Z}_-^{2j*}{\cal Z}_+^{1i*}
+c_\beta{\cal Z}_-^{1j*}{\cal Z}_+^{2i*}\Big)\Big]
\;,\nonumber\\
&&C_6^{A^0}(\mu_{_{\rm W}})=-C_5^{A^0}(\mu_{_{\rm W}})\;.\nonumber\\
\label{cpohp}
\end{eqnarray}
The box diagram contributions to the coefficients are
\begin{eqnarray}
&&C_1^{box}(\mu_{_{\rm W}})={1\over 4c_{_{\rm W}}^4}
{\bf G}_{_{L^{(a)}}}^{^{\{iiI\beta J\alpha\}}}
{\bf G}_{_{L^{(a)}}}^{^{\{jjJ\alpha J\beta\}}}
{\cal D}_{_{[111]}}^{^{(1)}}(
x_{_{\kappa_i^0}},x_{_{\kappa_j^0}},x_{_{\tilde{E}^\alpha}},
x_{_{\tilde{E}^\beta}})
\nonumber\\&&\hspace{2.2cm}+{\bf G}_{_{\nu^{(a)}}}^{^{\{iiI\beta J\beta\}}}
{\bf G}_{_{\nu^{(a)}}}^{^{\{jjJ\alpha J\alpha\}}}
{\cal D}_{_{[111]}}^{^{(1)}}(x_{_{\kappa_i^-}},
x_{_{\kappa_j^-}},x_{_{\tilde{\nu}^\alpha}},x_{_{\tilde{\nu}^\beta}})
\;,\nonumber\\
&&C_2^{box}(\mu_{_{\rm W}})={1\over 4c_{_{\rm W}}^4}
{\bf G}_{_{L^{(d)}}}^{^{\{iiI\beta J\alpha\}}}
{\bf G}_{_{L^{(d)}}}^{^{\{jjJ\alpha J\beta\}}}
{\cal D}_{_{[111]}}^{^{(1)}}(x_{_{\kappa_i^0}},x_{_{\kappa_j^0}}
,x_{_{\tilde{E}^\alpha}},x_{_{\tilde{E}^\beta}})
\nonumber\\&&\hspace{2.2cm}
+{1\over 4c_\beta^4}{\bf G}_{_{\nu^{(d)}}}^{^{\{iiI\beta J\beta\}}}
{\bf G}_{_{\nu^{(a)}}}^{^{\{jjJ\alpha J\alpha\}}}
x_{_{e^I}}^{1\over 2}x_{_{e^J}}^{1\over 3}
{\cal D}_{_{[111]}}^{^{(1)}}(x_{_{\kappa_i^-}},x_{_{\kappa_j^-}}
,x_{_{\tilde{\nu}^\alpha}},x_{_{\tilde{\nu}^\beta}})\;,\nonumber\\
&&C_3^{box}(\mu_{_{\rm W}})={1\over c_{_{\rm
W}}^4}{\bf G}_{_{L^{(b)}}}^{^{\{iiI\beta J\alpha\}}}
{\bf G}_{_{L^{(b)}}}^{^{\{jjJ\alpha J\beta\}}}
(x_{_{\kappa_i^0}}x_{_{\kappa_j^0}})^{1\over 2}
{\cal D}_{_{[111]}}^{^{(0)}}(x_{_{\kappa_i^0}},x_{_{\kappa_j^0}}
,x_{_{\tilde{E}^\alpha}},x_{_{\tilde{E}^\beta}})
\nonumber\\&&\hspace{2.2cm}
+{2\over c_\beta^2}{\bf G}_{_{\nu^{(b)}}}^{^{\{iiI\beta J\beta\}}}
{\bf G}_{_{\nu^{(b)}}}^{^{\{jjJ\alpha J\alpha\}}}
x_{_{e^J}}(x_{_{\kappa_i^-}}x_{_{\kappa_j^-}})^{1\over 2}
{\cal D}_{_{[111]}}^{^{(0)}}(x_{_{\kappa_i^-}},x_{_{\kappa_j^-}}
,x_{_{\tilde{\nu}^\alpha}},x_{_{\tilde{\nu}^\beta}})\;,
\nonumber\\
&&C_4^{box}(\mu_{_{\rm W}})={1\over c_{_{\rm W}}^4}\Big\{
{\bf G}_{_{L^{(b)}}}^{^{\{iiI\beta J\alpha\}}}
{\bf G}_{_{L^{(c)}}}^{^{\{iiJ\alpha J\beta\}}}(x_{_{\kappa_i^0}}
x_{_{\kappa_j^0}})^{1\over 2}{\cal D}_{_{[111]}}^{^{(0)}}(
x_{_{\kappa_i^0}},x_{_{\kappa_j^0}},x_{_{\tilde{E}^\alpha}},
x_{_{\tilde{E}^\beta}})
\nonumber\\&&\hspace{2.2cm}-{1\over 2}
{\bf G}_{_{L^{(a)}}}^{^{\{iiI\beta J\alpha\}}}
{\bf G}_{_{L^{(d)}}}^{^{\{jjJ\alpha J\beta\}}}
{\cal D}_{_{[111]}}^{^{(1)}}(
x_{_{\kappa_i^0}},x_{_{\kappa_j^0}},x_{_{\tilde{E}^\alpha}},
x_{_{\tilde{E}^\beta}})\Big\}\nonumber\\&&\hspace{2.2cm}
+{2\over c_\beta^2}
\Big\{{\bf G}_{_{\nu^{(a)}}}^{^{\{jiI\beta J\beta\}}}
{\bf G}_{_{\nu^{(d)}}}^{^{\{jiJ\alpha J\alpha\}}}
x_{_{e^J}}(x_{_{\kappa_i^-}}x_{_{\kappa_j^-}})^{1\over 2}
{\cal D}_{_{[111]}}^{^{(0)}}(x_{_{\kappa_i^-}},x_{_{\kappa_j^-}}
,x_{_{\tilde{\nu}^\alpha}},x_{_{\tilde{\nu}^\beta}})
\nonumber\\&&\hspace{2.2cm}
-{\bf G}_{_{\nu^{(a)}}}^{^{\{iiI\beta J\beta\}}}
{\bf G}_{_{\nu^{(d)}}}^{^{\{jjJ\alpha J\alpha\}}}{x_{_{e^J}}\over
2}{\cal D}_{_{[111]}}^{^{(1)}}(x_{_{\kappa_i^-}},x_{_{\kappa_j^-}}
,x_{_{\tilde{\nu}^\alpha}},x_{_{\tilde{\nu}^\beta}})\Big\}\;,
\nonumber\\
&&C_5^{box}(\mu_{_{\rm W}})={1\over c_{_{\rm W}}^4}\Big\{
{\bf G}_{_{L^{(a)}}}^{^{\{ijJ\alpha J\alpha\}}}
{\bf G}_{_{L^{(d)}}}^{^{\{jiI\beta J\beta\}}}
(x_{_{\kappa_i^0}}x_{_{\kappa_j^0}})
^{1\over 2}{\cal D}_{_{[111]}}^{^{(0)}}(x_{_{\kappa_i^0}},
x_{_{\kappa_j^0}},x_{_{\tilde{E}^\alpha}},x_{_{\tilde{E}^\beta}})
\nonumber\\&&\hspace{2.2cm}-{1\over 2}
{\bf G}_{_{L^{(d)}}}^{^{\{iiI\beta J\alpha\}}}
{\bf G}_{_{L^{(a)}}}^{^{\{jjJ\alpha J\beta\}}}
{\cal D}_{_{[111]}}^{^{(1)}}(x_{_{\kappa_i^0}},x_{_{\kappa_j^0}}
,x_{_{\tilde{E}^\alpha}},x_{_{\tilde{E}^\beta}})
\Big\}\nonumber\\&&\hspace{2.2cm}
+{2\over c_\beta^2}\Big\{{\bf G}_{_{\nu^{(a)}}}^{^{\{ijI\beta J\beta\}}}
{\bf G}_{_{\nu^{(d)}}}^{^{\{ijJ\alpha J\alpha\}}}
(x_{_{e^I}}x_{_{e^J}}x_{_{\kappa_i^-}}
x_{_{\kappa_j^-}})^{1\over 2}{\cal D}_{_{[111]}}^{^{(0)}}(
x_{_{\kappa_i^-}},x_{_{\kappa_j^-}},x_{_{\tilde{\nu}^\alpha}}
,x_{_{\tilde{\nu}^\beta}})\nonumber\\&&\hspace{2.2cm}
-{\bf G}_{_{\nu^{(a)}}}^{^{\{jjI\beta J\beta\}}}
{\bf G}_{_{\nu^{(d)}}}^{^{\{iiJ\alpha J\alpha\}}}{(x_{_{e^I}}x_{_{e^J}}
)^{1\over 2}\over 2}{\cal D}_{_{[111]}}^{^{(1)}}(x_{_{\kappa_i^-}}
,x_{_{\kappa_j^-}},x_{_{\tilde{\nu}^\alpha}},x_{_{\tilde{\nu}^\beta}})
\Big\}\;,
\nonumber\\
&&C_6^{box}(\mu_{_{\rm W}})={1\over c_{_{\rm
W}}^4}{\bf G}_{_{L^{(c)}}}^{^{\{iiI\beta J\alpha\}}}
{\bf G}_{_{L^{(c)}}}^{^{\{jjJ\alpha J\beta\}}}
(x_{_{\kappa_i^0}}x_{_{\kappa_j^0}})^{1\over 2}
{\cal D}_{_{[111]}}^{^{(0)}}(x_{_{\kappa_i^0}},x_{_{\kappa_j^0}}
,x_{_{\tilde{E}^\alpha}},x_{_{\tilde{E}^\beta}})
\nonumber\\&&\hspace{2.2cm}
+{2\over c_\beta^2}
{\bf G}_{_{\nu^{(a)}}}^{^{\{ijI\beta J\beta\}}}
{\bf G}_{_{\nu^{(d)}}}^{^{\{jiJ\alpha J\alpha\}}}
(x_{_{e^I}}x_{_{e^J}}x_{_{\kappa_i^-}}x_{_{\kappa_j^-}})^{1\over 2}
{\cal D}_{_{[111]}}^{^{(0)}}(x_{_{\kappa_i^-}},x_{_{\kappa_j^-}},
x_{_{\tilde{\nu}^\alpha}},x_{_{\tilde{\nu}^\beta}})\;.
\nonumber\\ \label{box}
\end{eqnarray}

\section{The loop integral functions \label{app4}}

In this appendix, we present the expressions of various loop
integral functions which appear in the text.
\begin{eqnarray}
&&f_1(x_1,x_2)=\left\{\begin{array}{l}
{x_1x_2^2(\ln x_1-\ln x_2)\over (x_2-x_1)^4}
-{x_1^2-5x_1x_2-2x_2^2\over 6(x_2-x_1)^3}\;,({\rm for}\;
\;x_1\neq x_2),\\ \\
{1\over 12x_1}\;,\;\;({\rm for}\;\;x_1=x_2);\\
\end{array}\right.
\nonumber\\
&&f_2(x_1,x_2)=\left\{\begin{array}{l}
{(2x_1x_2-x_2^2)\ln x_2-x_1^2\ln x_1\over 2(x_2-x_1)^2}
-{x_1\over 2(x_2-x_1)}-{x_1\over 2(x_2-x_1)}\;,\;\; ({\rm for}
\;\;x_1\neq x_2),\\ \\
-{1\over 4}-{1\over 2}\ln x_1\;,\;\;({\rm for}
\;\;x_1= x_2);
\end{array}\right.\nonumber\\
&&B_{_{[21]}}^{^{(0)}}(x_1,x_2)=\left\{\begin{array}{l}
{x_2(\ln x_1-\ln x_2)\over (x_2-x_1)^2}+{1\over x_2-x_1}\;,
({\rm for}\;\;x_1\neq x_2),\\ \\
-{1\over 2x_1}\;,\;\;({\rm for}\;\;x_1= x_2);
\end{array}\right.\nonumber\\
&&B_{_{[31]}}^{^{(0)}}(x_1,x_2)=\left\{\begin{array}{l}
{x_2(\ln x_1-\ln x_2)\over (x_2-x_1)^3}+{x_1+x_2\over 2x_1(x_2-x_1)^2}\;,
({\rm for}\;\;x_1\neq x_2),\\ \\
{1\over 6x_1^2}\;,\;\;({\rm for}\;\;x_1= x_2);
\end{array}\right.\nonumber\\
&&B_{_{[41]}}^{^{(0)}}(x_1,x_2)=\left\{\begin{array}{l}
{x_2(\ln x_1-\ln x_2)\over (x_2-x_1)^4}+{2x_1^2+5x_1x_2
-x_2^2\over 6x_1^2(x_2-x_1)^3}\;,
({\rm for}\;\;x_1\neq x_2),\\ \\
-{1\over 12x_1^3}\;,\;\;({\rm for}\;\;x_1= x_2);
\end{array}\right.\nonumber\\
&&B_{_{[31]}}^{^{(1)}}(x_1,x_2)=B_{_{[21]}}^{^{(0)}}(x_1,x_2)
+x_1B_{_{[31]}}^{^{(0)}}(x_1,x_2)\;,\nonumber\\
&&B_{_{[41]}}^{^{(2)}}(x_1,x_2)=B_{_{[21]}}^{^{(0)}}(x_1,x_2)
+2x_1B_{_{[31]}}^{^{(0)}}(x_1,x_2)+x_1^2B_{_{[41]}}^{^{(0)}}(x_1,x_2)
\;,\nonumber\\
&&{\cal T}_{_{[111]}}^{^{(0)}}(x_1,x_2,x_3)=\left\{\begin{array}{l}
-{x_1\ln x_1\over (x_2-x_1)(x_3-x_1)}-{x_2\ln x_2\over (x_1-x_2)(x_3-x_2)}
-{x_3\ln x_3\over (x_1-x_3)(x_2-x_3)}\;,
({\rm for}\;\;x_1\neq x_2\neq x_3),\\ \\
{x_3(\ln x_1-\ln x_3)\over (x_3-x_1)^2}+{1\over x_3-x_1}\;,
({\rm for}\;\;x_1= x_2\neq x_3),\\ \\
-{1\over 2x_1}\;,\;\;({\rm for}\;\;x_1= x_2=x_3);
\end{array}\right.\nonumber\\
&&{\cal T}_{_{[111]}}^{^{(1)}}(x_1,x_2,x_3)=\left\{\begin{array}{l}
-{x_1^2\ln x_1\over (x_2-x_1)(x_3-x_1)}-{x_2^2\ln x_2\over (x_1-x_2)(x_3-x_2)}
-{x_3^2\ln x_3\over (x_1-x_3)(x_2-x_3)}\;,
({\rm for}\;\;x_1\neq x_2\neq x_3),\\ \\
{(2x_1x_3-x_1^2)\ln x_1-x_3^2\ln x_3\over (x_3-x_1)^2}+{x_1\over x_3-x_1}\;,
({\rm for}\;\;x_1= x_2\neq x_3),\\ \\
-{3\over 2}-\ln x_1\;,\;\;({\rm for}\;\;x_1= x_2=x_3);
\end{array}\right.\nonumber\\
&&{\cal D}_{_{[1111]}}^{^{(0)}}(x_1,x_2,x_3,x_4)=\left\{\begin{array}{l}
-{x_1\ln x_1\over (x_2-x_1)(x_3-x_1)(x_4-x_1)}-{x_2\ln x_2\over (x_1-x_2)
(x_3-x_2)(x_4-x_2)}\\-{x_3\ln x_3\over
(x_1-x_3)(x_2-x_3)(x_4-x_3)}
-{x_4\ln x_4\over (x_1-x_4)(x_2-x_4)(x_3-x_4)}\;,\;\;
({\rm for}\;\;x_1\neq x_2\neq x_3\neq x_4),\\ \\
{x_3\ln x_3\over (x_1-x_3)^2(x_4-x_3)}+{x_4\ln x_4
\over (x_1-x_4)^2(x_3-x_4)}\\
-{(x_3x_4-x_1^2)
\ln x_1\over (x_3-x_1)^2(x_4-x_1)^2}-{1\over (x_3-x_1)(x_4-x_1)}
\;,({\rm for}\;\;x_1=x_2\neq x_3\neq x_4),\\ \\
-{2x_1\ln x_1-2x_3\ln x_3\over (x_3-x_1)^3} -{2+\ln x_1+\ln
x_3\over (x_3-x_1)^2}\;,
\;\;({\rm for}\;\;x_1= x_2\neq x_3=x_4);\\ \\
{x_4(\ln x_1-\ln x_4)\over (x_4-x_1)^3}+{x_4\over
2x_1(x_4-x_1)^2}\;,
({\rm for}\;\;x_1= x_2=x_3\neq x_4),\\ \\
{1\over 6x_1^2}\;,\;\;({\rm for}\;\;x_1= x_2=x_3=x_4);
\end{array}\right.\nonumber\\
&&{\cal D}_{_{[1111]}}^{^{(1)}}(x_1,x_2,x_3,x_4)=\left\{\begin{array}{l}
-{x_1^2\ln x_1\over (x_2-x_1)(x_3-x_1)(x_4-x_1)}-{x_2^2\ln x_2\over (x_1-x_2)
(x_3-x_2)(x_4-x_2)}\\-{x_3^2\ln x_3\over
(x_1-x_3)(x_2-x_3)(x_4-x_3)}
-{x_4^2\ln x_4\over (x_1-x_4)(x_2-x_4)(x_3-x_4)}\;,\;\;
({\rm for}\;\;x_1\neq x_2\neq x_3\neq x_4),\\ \\
{x_3^2\ln x_3\over (x_1-x_3)^2(x_4-x_3)}+{x_4^2\ln x_4
\over (x_1-x_4)^2(x_3-x_4)}\\
-{(2x_1x_3x_4-x_1^2x_3-x_1^2x_4)
\ln x_1\over (x_3-x_1)^2(x_4-x_1)^2}-{x_1\over (x_3-x_1)(x_4-x_1)}
\;,({\rm for}\;\;x_1=x_2\neq x_3\neq x_4),\\ \\
-{2x_1^2\ln x_1-2x_3^2\ln x_3\over (x_3-x_1)^3}
-{(x_1+x_3+2x_1\ln x_1+2x_3\ln x_3)\over (x_3-x_1)^2}\;,
\;\;({\rm for}\;\;x_1= x_2\neq x_3=x_4);\\ \\
{x_4^2(\ln x_1-\ln x_4)\over (x_4-x_1)^3}+{3x_4-x_1\over 2(x_4-x_1)^2}\;,
({\rm for}\;\;x_1= x_2=x_3\neq x_4),\\ \\
-{1\over 3x_1}\;,\;\;({\rm for}\;\;x_1= x_2=x_3=x_4).
\end{array}\right.
\label{loopfun}
\end{eqnarray}
\section{The expression for the $\lambda_i$ parameters\label{apd2}}

The $\lambda$ parameters that involve the neutral Higgs mass
mixing are given as
\begin{eqnarray}
\lambda_1 \!\!&=&\!\!
   -\, \frac{g_1^2+g_2^2}{8}\  \Big( 1\, -\, \frac{3}{8\pi^2}\
          h^2_b\, t \Big)\nonumber\\
   &&-\,\frac{3}{16\pi^2}\ h^4_b\, \Big[\, t\, +\, \frac{1}{2}\, X_b\, +\,
   \frac{1}{16\pi^2}\, \Big(\, \frac{3}{2}\, h^2_b\, +\, \frac{1}{2}\,
   h^2_t\, -\, 8g^2_s\, \Big)\, (X_b t\, +\, t^2 )\, \Big]\nonumber\\
   &&+\, \frac{3}{192\pi^2}\, h^4_t\, \frac{|\mu|^4}{M^4_{\rm SUSY}}\,
   \Big[\, 1\, +\, \frac{1}{16\pi^2}\, (9h^2_t\, -\, 5h^2_b\, -\,
   16g^2_s)t\, \Big] ,\nonumber\\
  \label{lambda2}
\lambda_2 \!\!&=&\!\!
   -\, \frac{g_1^2+g_2^2}{8}\  \Big( 1\, -\, \frac{3}{8\pi^2}\
          h^2_t\, t \Big)\nonumber\\
   &&-\,\frac{3}{16\pi^2}\ h^4_t\, \Big[\, t\, +\, \frac{1}{2}\, X_t\, +\,
   \frac{1}{16\pi^2}\, \Big(\, \frac{3}{2}\, h^2_t\, +\, \frac{1}{2}\,
   h^2_b\, -\, 8g^2_s\, \Big)\, (X_t t\, +\, t^2 )\, \Big]\nonumber\\
   &&+\, \frac{3}{192\pi^2}\, h^4_b\, \frac{|\mu|^4}{M^4_{\rm SUSY}}\,
   \Big[\, 1\, +\, \frac{1}{16\pi^2}\, (9h^2_b\, -\, 5h^2_t\, -\,
   16g^2_s)t\, \Big] ,\nonumber\\
\lambda_3 \!\!&=&\!\!
   -\, \frac{g_2^2-g_1^2}{4}\  \Big[\, 1\, -\, \frac{3}{16\pi^2}\
          (h^2_t\, +\, h^2_b) \, t\, \Big]\nonumber\\
   &&-\,\frac{3}{8\pi^2}\ h^2_th^2_b\, \Big[\, t\, +\, \frac{1}{2}\,
          X_{tb}\, +\, \frac{1}{16\pi^2}\,
   ( h^2_t\, +\, h^2_b\, -\, 8g^2_s)\, (X_{tb} t\, +\, t^2 )\, \Big]\nonumber\\
   &&-\, \frac{3}{96\pi^2}\, h^4_t\, \Big(\, \frac{3|\mu|^2}{M^2_{\rm
   SUSY}}\, -\, \frac{|\mu|^2 |A_t|^2}{M^4_{\rm SUSY}}\, \Big)\,
   \Big[\, 1\, +\, \frac{1}{16\pi^2}\, (6h^2_t\, -\, 2h^2_b\, -\,
   16g^2_s)t\, \Big]\nonumber\\
   &&-\, \frac{3}{96\pi^2}\, h^4_b\, \Big(\, \frac{3|\mu|^2}{M^2_{\rm
   SUSY}}\, -\, \frac{|\mu|^2 |A_b|^2}{M^4_{\rm SUSY}}\, \Big)\,
   \Big[\, 1\, +\, \frac{1}{16\pi^2}\, (6h^2_b\, -\, 2h^2_t\, -\,
   16g^2_s)t\, \Big] ,\nonumber\\
\lambda_4 \!\!&=&\!\! \frac{g_2^2}{2}\  \Big[\, 1\, -\, \frac{3}{16\pi^2}\
          (h^2_t\, +\, h^2_b) \, t\, \Big]\nonumber\\
   &&+\,\frac{3}{8\pi^2}\ h^2_th^2_b\, \Big[\, t\, +\, \frac{1}{2}\,
          X_{tb}\, +\, \frac{1}{16\pi^2}\,
   ( h^2_t\, +\, h^2_b\, -\, 8g^2_s)\, (X_{tb} t\, +\, t^2 )\, \Big]\nonumber\\
   &&-\, \frac{3}{96\pi^2}\, h^4_t\, \Big(\, \frac{3|\mu|^2}{M^2_{\rm
   SUSY}}\, -\, \frac{|\mu|^2 |A_t|^2}{M^4_{\rm SUSY}}\, \Big)\,
   \Big[\, 1\, +\, \frac{1}{16\pi^2}\, (6h^2_t\, -\, 2h^2_b\, -\,
   16g^2_s)t\, \Big]\nonumber\\
   &&-\, \frac{3}{96\pi^2}\, h^4_b\, \Big(\, \frac{3|\mu|^2}{M^2_{\rm
   SUSY}}\, -\, \frac{|\mu|^2 |A_b|^2}{M^4_{\rm SUSY}}\, \Big)\,
   \Big[\, 1\, +\, \frac{1}{16\pi^2}\, (6h^2_b\, -\, 2h^2_t\, -\,
   16g^2_s)t\, \Big] ,\nonumber\\
\lambda_5 \!\!&=&\!\! \frac{3}{192\pi^2}\, h^4_t\,
   \frac{\mu^2 A_t^2}{M^4_{\rm SUSY}}\,
   \Big[\, 1\, -\, \frac{1}{16\pi^2}\, (2h^2_b\, -\, 6h^2_t\, +\,
   16g^2_s)t\, \Big]\nonumber\\
   &&+\, \frac{3}{192\pi^2}\, h^4_b\,
   \frac{\mu^2 A_b^2}{M^4_{\rm SUSY}}\,
   \Big[\, 1\, -\, \frac{1}{16\pi^2}\, (2h^2_t\, -\, 6h^2_b\, +\,
   16g^2_s)t\, \Big] ,\nonumber\\
\lambda_6 \!\!&=&\!\! -\, \frac{3}{96\pi^2}\, h^4_t\,
   \frac{|\mu|^2 \mu A_t}{M^4_{\rm SUSY}}\,
   \Big[\, 1\, -\, \frac{1}{16\pi^2}\, \Big(\, \frac{7}{2}\, h^2_b\,
   -\, \frac{15}{2}\, h^2_t\, +\, 16g^2_s\, \Big)\, t\, \Big]\nonumber\\
   &&\!\!\!+\, \frac{3}{96\pi^2}\, h^4_b\, \frac{\mu}{M_{\rm SUSY}}\,
   \Big(\, \frac{6A_b}{M_{\rm SUSY}}\, -\, \frac{|A_b|^2 A_b}{M^3_{\rm
   SUSY}}\, \Big)
  \Big[\, 1\, -\, \frac{1}{16\pi^2}\, \Big(\, \frac{1}{2}\,
   h^2_t\, -\, \frac{9}{2}\, h^2_b\, +\, 16g^2_s\, \Big)\, t\, \Big], \qquad
\nonumber\\
\lambda_7 \!\!&=&\!\! -\, \frac{3}{96\pi^2}\, h^4_b\,
   \frac{|\mu|^2 \mu A_b}{M^4_{\rm SUSY}}\,
   \Big[\, 1\, -\, \frac{1}{16\pi^2}\, \Big(\, \frac{7}{2}\, h^2_t\,
   -\, \frac{15}{2}\, h^2_b\, +\, 16g^2_s\, \Big)\, t\, \Big]\nonumber\\
   &&\!\!\!+\, \frac{3}{96\pi^2}\, h^4_t\, \frac{\mu}{M_{\rm SUSY}}\,
   \Big(\, \frac{6A_t}{M_{\rm SUSY}}\, -\,
   \frac{|A_t|^2 A_t}{M^3_{\rm SUSY}}\,\Big)
   \Big[\, 1\, -\, \frac{1}{16\pi^2}\, \Big(\, \frac{1}{2}\,
   h^2_b\, -\, \frac{9}{2}\, h^2_t\, +\, 16g^2_s\, \Big)\, t\, \Big] ,
\label{lamb}
\end{eqnarray}
where $t = \ln ( M^2_{\rm SUSY}/\overline{m}^2_t ),\;g_1={e\over c_{_{\rm W}}}
,\;g_2={e\over s_{_{\rm W}}}$ and
\begin{eqnarray}
  \label{hthb}
h_t &=& \frac{\sqrt{2}\, m_t (\overline{m}_t)}{v \sin\beta }\ ,\, \qquad
h_b \ \, =\ \, \frac{\sqrt{2}\, m_b (\overline{m}_t)}{v \cos\beta }\ ,\nonumber\\
  \label{Xtbtb}
X_t &=& \frac{2|A_t|^2}{M^2_{\rm SUSY}}\ \Big(\, 1\, -\,
     \frac{|A_t|^2}{12M^2_{\rm SUSY}}\, \Big)\, ,\nonumber\\
X_b &=& \frac{2|A_b|^2}{M^2_{\rm SUSY}}\ \Big(\, 1\, -\,
     \frac{|A_b|^2}{12M^2_{\rm SUSY}}\, \Big)\, , \nonumber\\
X_{tb} &=& \frac{ |A_t|^2+|A_b|^2 + 2{\rm Re}(A^*_bA_t) }
     {2\,M^2_{\rm SUSY}}\, -\, \frac{|\mu|^2}{M^2_{\rm SUSY}}\,
     -\, \frac{|\,|\mu|^2 -A^*_bA_t\,|^2}{6\,M^4_{\rm SUSY}}\ .\quad
\end{eqnarray}
In Eq. (\ref{hthb}), $\overline{m}_t$ is the top-quark pole mass, which
is related to the on-shell running mass $m_t$ through
\begin{equation}
  \label{mt}
m_t (\overline{m}_t)\ =\ \frac{\overline{m}_t}{1 + \frac{4}{3\pi}\,
 \alpha_s (\overline{m}_t )}\ .
\end{equation}


\begin{figure}
\setlength{\unitlength}{1mm}
\begin{center}
\begin{picture}(230,200)(55,90)
\put(50,80){\includegraphics{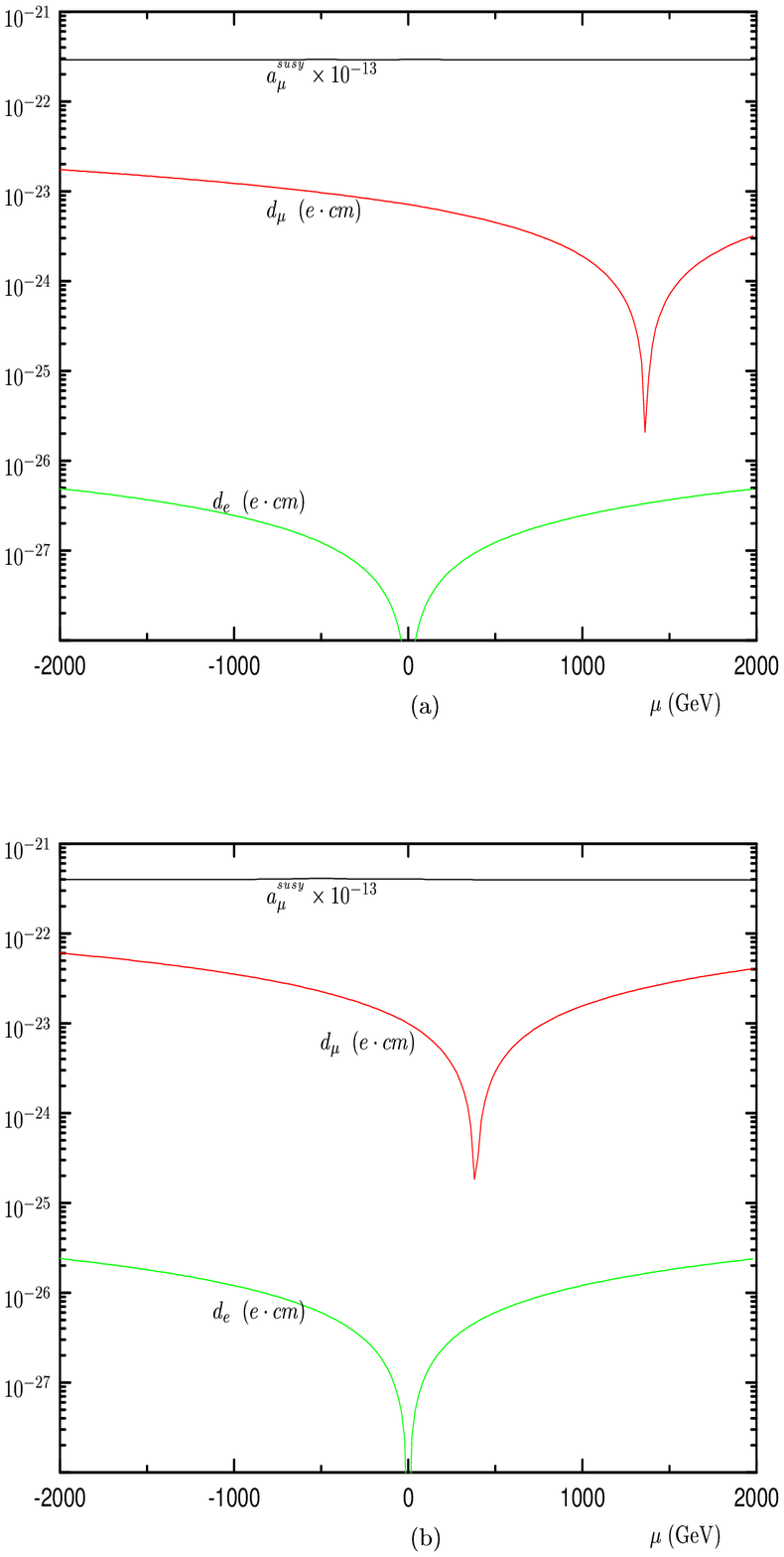}}
\end{picture}
\caption[]{The muon's MDM and EDM, the electron's EDM verus the
parameter $\mu$ in MSSM with (a)$\tan\beta=5$, (b)$\tan\beta=50$;
the other parameters are the same as in the text.} \label{fig6}
\end{center}
\end{figure}
\begin{figure}
\setlength{\unitlength}{1mm}
\begin{center}
\begin{picture}(230,200)(55,90)
\put(50,80){\includegraphics{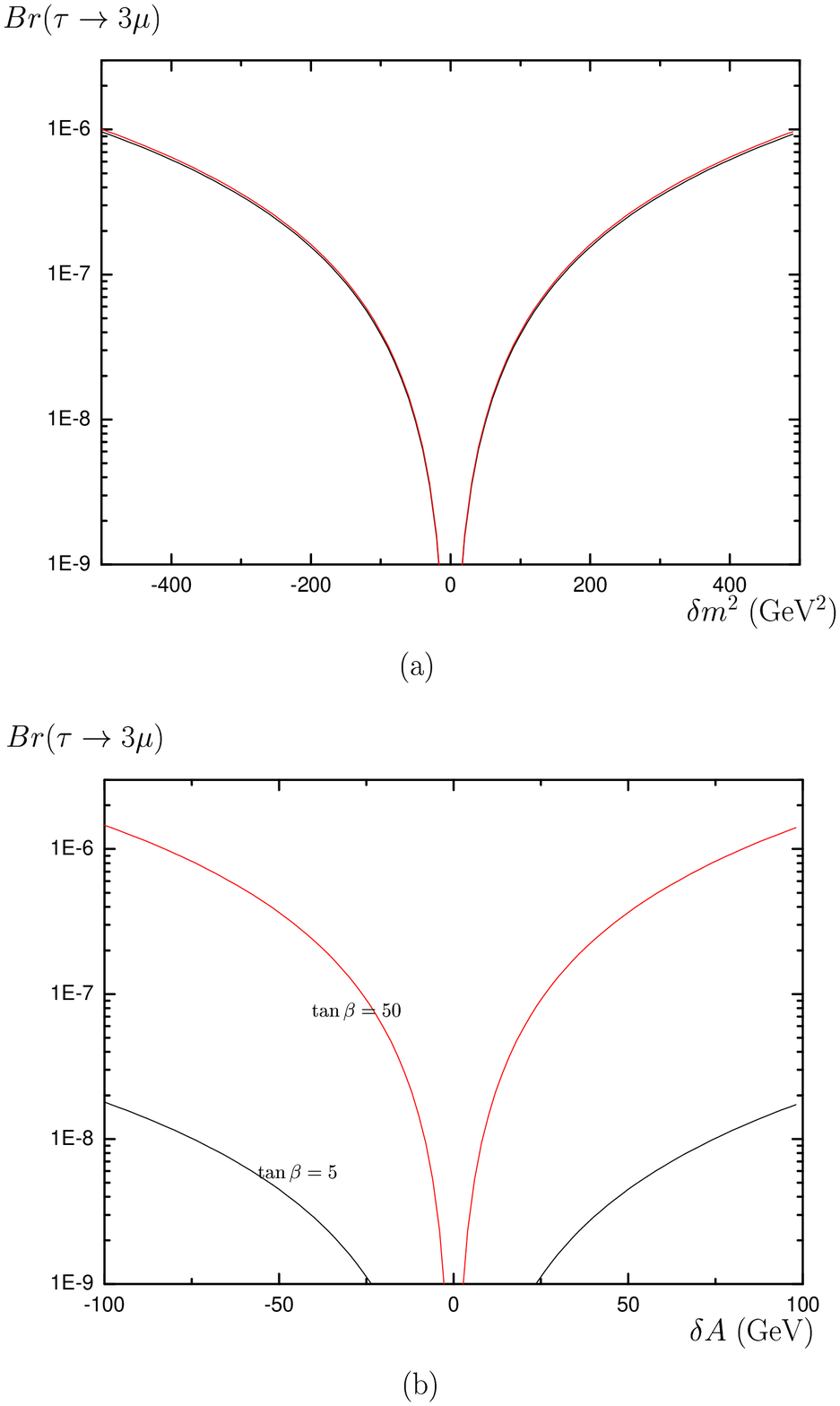}}
\end{picture}
\caption[]{With $\mu=200\;{\rm GeV}$ and all zero $CP$ phases, the
branching ratio of $\tau\rightarrow 3\mu$ versus (a)
$m_{_{L_{23}}}^2=m_{_{R_{23}}}^2=\delta m^2$ or (b)
$A^l_{23}=A^l_{23}=\delta A$, the other parameters are taken as in
the text.} \label{fig7}
\end{center}
\end{figure}
\begin{figure}
\setlength{\unitlength}{1mm}
\begin{center}
\begin{picture}(230,200)(55,90)
\put(50,80){\includegraphics{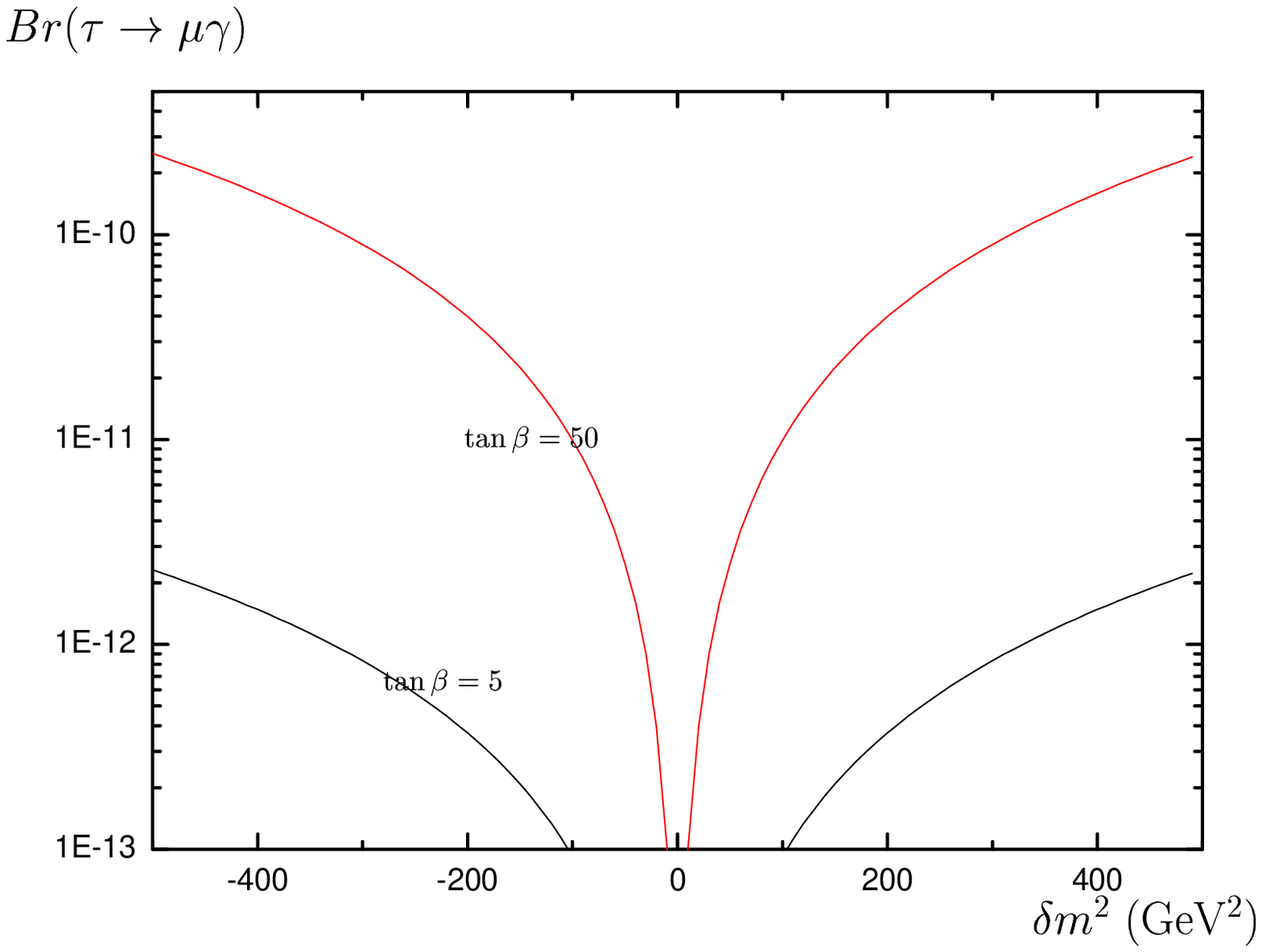}}
\end{picture}
\caption[]{The the $BR(\tau\rightarrow\mu\gamma)$ versus
$m_{_{L_{23}}}^2=m_{_{R_{23}}}^2=\delta m^2$, the other parameters
are taken as in the text.} \label{fig8}
\end{center}
\end{figure}

\begin{thebibliography}{99}
\bibitem{Bennett}G. W. Bennett {\it et al.}, Phys. Rev. Lett.
{\bf 89}, 101804(2002)(hep-ex/0208001).
\bibitem{Wise}M. Ramsey-Musolf and M. B. Wise, Phys. Rev. Lett.
{\bf 89}, 041601(2002).
\bibitem{Komine}S. Komine, T. Moroi and M. Yamaguchi,
Phy. Lett. {\bf B506}, 93(2001) (hep-ph/0102204).
\bibitem{Mahanta}U. Mahanta, Phys. Lett. {\bf B511}, 235(2001)
(hep-ph/0102211).
\bibitem{Kinoshita}T. Kinoshita, B. Nizic, and Y. Okamoto,
Phys. Rev. {\bf D41}, 593(1990); T. Kinoshita, Phys. Rev.
{\bf D47}, 5013(1993); M. Samuel and G. Li, Phys. Rev.
{\bf D44}, 3935(1991); 48, 1879(E)(1993); S. Laporta
and E. Remiddi, Phys. Lett. {\bf B312}, 495(1993);
S. Karshenboim, Yad. Fiz. {\bf 56}, 252(1993).
\bibitem{Kinoshita1}T. Kinoshita, Phys. Rev. Lett.
{\bf 75},4728(1995).
\bibitem{Davier}M. Davier and A. H\"{o}cker, hep-ph/9801361;
B. Krause, Phys. Lett. {\bf B390}, 392(1997); M. Hayakawa and T.
Kinoshita, Phys. Rev. {\bf D57}, 465(1998); S. Narison, Phys.
Lett. {\bf B513}, 53(2001) (hep-ph/0103199); K. Hagiwara, A. D.
Martin, D. Nomura, T. Teubner, hep-ph/0209187; S. I. Eidelman, S.
G. Karshenboim, V. A. Shelyuto, hep-ph/0209146.
\bibitem{Fujikawa}K. Fujikawa, B. W. Lee and A. I. Sanda,
Phys. Rev. {\bf D6}, 2923(1972).
\bibitem{Jackiw}R. Jackiw and S. Weinberg, Phys. Rev. {\bf D5},
2473(1972).
\bibitem{Altarelli}G. Altarelli, N. Cabibbo and L. Maiani,
Phys. Lett. {\bf B40}, 415(1972).
\bibitem{Bars}I. Bars and M. Yoshimura, Phys. Rev. {\bf D6},
374(1972).
\bibitem{Bardeen}W. A. Bardeen, R. Gastmans and B. E. Lautrup,
Nucl. Phys. {\bf B46}, 315(1972).
\bibitem{Czarnecki}A. Czarnecki, B. Krause, and W. J. Marciano,
Phys. Rev. Lett. {\bf 76}, 3267(1996); G. Degrassi and G. F.
Giudice, Phys. Rev. {\bf D58}, 053007(1998).
\bibitem{Moroi}T. Moroi, Phys. Rev. {\bf D53}, 6565(1996);
U. Chattopadhyay and P. Nath, Phys. Rev. {\bf D53}, 1648(1996); M.
Byran, C. Kolda and J. E. Lennon hep-ph/0208067.
\bibitem{Ibrahim}T. Ibrahim and P. Nath, Phys. Rev. {\bf D61},
095008(2000); {\it ibid.}{\bf 62}, 015004,2000.
\bibitem{Hisano}J. Hisano and D. Nomura, Phys. Rev. {\bf D59},
116005(1999); J. Hisano, T. Moroi, K. Tobe and M. Yamaguchi, Phys.
Rev. {\bf D53}, 2442(1996); Phys. Lett. {\bf B357}, 579(1995);
{\it ibid.} {\bf B391}, 341(1997); J. Hisano, D. Nomura, Y. Okada,
Y. Shimizu and T. Tanaka, Phys. Rev. {\bf D58}, 116010(1998); J.
Hisano, D. Nomura and Y. Yanagida, Phys. Lett. {\bf B437},
351(1998); J. Hisano and K. Tobe, Phys. Lett. {\bf B510},
197(2001) (hep-ph/0102315).
\bibitem{Kim}J. Kim, B. Kyae and H. Lee, Phys. Lett. {\bf B520},
298(2001) (hep-ph/0103054).
\bibitem{Lopez}J. Lopez, D. Nanopoulos and X. Wang,
Phys. Rev. {\bf D49}366, (1994).
\bibitem{Carena}M. Carena, G. F. Giudice and C. E. M. Wagner,
Phys. Lett. {\bf B390}, 234(1997); J. L. Feng and K. T. Matchev,
Phys. Rev. Lett. {\bf 86}, 3480(2001) (hep-ph/0102146); J. L.
Feng, K. T. Matchev and Y. Shadmi, hep-ph/0208106; L. Everett, G.
L. Kane, S. Rigolin, L. Wang, Phys. Rev. Lett. {\bf 86},
3484(2001) (hep-ph/0102145); E. Ma, M. Raidal, Phys. Rev. Lett.
{\bf 87}, 011802(2001) (hep-ph/0102255); S. P. Martin and J. D.
Wells, Phys. Rev. {\bf D64}, 035003(2001) (hep-ph/0103067),
hep-ph/0209309; E. Baltz, P. Gondolo, Phys. Rev. Lett. {\bf 86},
5004(2001) (hep-ph/0102147); U. Chattopadhyay and P. Nath, Phys.
Rev. Lett. {\bf 86}, 5854(2001) (hep-ph/0102157); S. Komine, T.
Moroi and M. Yamaguchi, Phys. Lett. {\bf B506}, 93(2001) (
hep-ph/0102204); T. Ibrahim, U. Chattopadhyay, P. Nath, Phys. Rev.
{\bf D64}, 016010(2001) (hep-ph/0102324); R. Arnowitt, B. Dutta,
B. Hu and Y. Santoso, Phys. Lett. {\bf B505}, 177(2001) (
hep-ph/0102344); T. Kobayashi, H. Terao, Phys. Rev. {\bf D64},
075003(2001) (hep-ph/0103028); K. Choi, K. Hwang, S. Kang, K. Lee,
W. Song, Phys. Rev. {\bf D64}, 055001(2001) (hep-ph/0103048); S.
Komine, T. Moroi, M. Yamaguchi, Phys. Lett. {\bf B507}, 224(2001)
(hep-ph/0103182); A. Bartl, T. Gajdosik, E. Lunghi, A. Masiero, W.
Porod, H. Stremnitzer, O. Vives, Phys. Rev. {\bf D64},
076009(2001) (hep-ph/0103324).
\bibitem{THDM}R. A. Diaz, R. Martinez, J.-Alexis Rodriguez,
Phys. Rev. {\bf D64}, 033004(2001) (hep-ph/0103050); S. K. Kang,
K. Y. Lee, P:hys. Lett. {\bf B521}, 61(2001) (hep-ph/0103064); C.
A. de S. Pires, P. S. Rodrigues da Silva, Phys. Rev. {\bf D64},
117701(2001) (hep-ph/0103083); E. O. Iltan, hep-ph/0103105; K.
Cheung, C.-H. Chou, O. C. W. Kong, Phys. Rev. {\bf D64},
111301(2001) (hep-ph/0103183); M. Krawczyk, hep-ph/0103223; F.
Larios, G. Tavares-Velasco, C.-P. Yuan, Phys. Rev. {\bf D64},
055004(2001) (hep-ph/0103292).
\bibitem{new}A. Czarnecki, W. J. Marciano, Phys. Rev. {\bf D64},
013014(2001) (hep-ph/0102122); U. Mahanta, Eur. Phys. J. {\bf
C21}, 171(2001) (hep-ph/0102176); D. Chakraverty, D. Choudhury, A.
Datta, Phys. Lett. {\bf B506}, 103(2001) (hep-ph/0102180);  S.
Gninenko, N. V. Krasnikov, Phys. Lett. {\bf B513}, 119(2001) (
hep-ph/0102222); K. Cheung, Phys. Rev. {\bf D64}, 033001(2001) (
hep-ph/0102238); P. das, S. K. Rai, S. Raychaudhuri,
hep-ph/0102242; T. Kephart, H. P\"{a}s, Phys. Rev. {\bf D65},
093014(2001) (hep-ph/0102243); Z. Xing, Phys. Rev. {\bf D64},
017304(2001) (hep-ph/0102304); X. Calmet, H. Fritzsch, D.
Holtmannspotter, Phys. Rev. {\bf D64}, 037701(2001) (
hep-ph/0103012; S. Rajpoot, hep-ph/0103069; S. C. Park, H. S.
Song, hep-ph/0103072); C. S. Kim, J. D. Kim, J. Song, Phys. Lett.
{\bf B511}, 251(2001) (hep-ph/0103127); E. Mitsuda, K. Sasaki,
hep-ph/0103202; S. Baek, P. Ko, H. S. Lee, Phys. Rev. {\bf D65},
035004(2001) (hep-ph/0103218).
\bibitem{Arzt}C. Arzt, M. B. Einhorn and J. Wudka, Phys. Rev.
{\bf D49}, 1370(1994); U. Mahanta, Phys. Lett. {\bf B511},
235(2001) (hep-ph/0102211); M. B. Einhorn, J. Wudka, Phys. Rev.
Lett. {\bf 87}, 071805(2001) (hep-ph/0103034); M. Raidal, Phys.
Lett. {\bf B508}, 51(2001) (hep-ph/0103224).
\bibitem{Huang}T. Huang, Z.-H. Lin, L.-Y. Shan and X. Zhang,
Phys. Rev. {\bf D64}, 071301(2001) (hep-ph/0102193).
\bibitem{Barr}S. M. Barr, A. Zee, Phys. Rev. Lett. {\bf 65},21(1990);
D. Chang, W.-Y. Keung, T. C. Yuan, Phys. Lett. {\bf B251},608(1990);
J. F. Gunion, D. Wyler, Phys. Lett. {\bf B248}, 170(1990); C. Kao,
R. -M. Xu, Phys. Lett. {\bf B296}, 435(1992).
\bibitem{Pilaftsis4}D. Chang, W.-Y. Keung, A. Pilaftsis, Phys. Rev. Lett.
{\bf 82}, 900(1999); Phys. Rev. Lett. {\bf 83}, 3972(1999)(E); A.Pilaftsis,
Phys. Lett. {\bf B471}, 174(1999); D. Chang, W. -F. Chang, W.-Y. Keung,
Phys. Lett. {\bf B478}, 239(2000); A.Pilaftsis, Nucl. Phys. {\bf B644}, 263(2002).
\bibitem{PDG}Particle Data Group, Eur. Phys. J. {\bf C15},
1(2000).
\bibitem{Pilaftsis1}A. Pilaftsis, Phys. Lett. {\bf B435},88(1998);
Phys. Rev. {\bf D58},096010(1998).
\bibitem{Pilaftsis2}A. Pilaftsis, Nucl. Phys. {\bf B504}, 61(1997).
\bibitem{Pilaftsis3}A. Pilaftsis, C. E. M. Wagner, Nucl. Phys. {\bf B553},
3{1999}; M. Carena, J. Ellis, A. Pilaftsis, C. E. M. Wagner, Nucl. Phys. {\bf B586},
92(2000); Nucl. Phys. {\bf B625}, 345(2002).
\bibitem{notation}J. Rosiek, Phys. Rev. {\bf D41},
3464(1990); Chao-Hsi Chang and Tai-Fu Feng, Eur.
Phys. J. {\bf C12}, 137(2000).
\bibitem{Gabbiani}F. Gabbiani and A. Masiero, Nucl. Phys. {\bf
B322}, 235(1989); F. Gabbiani, E. Gabrielli, A. Masiero, and
L. Silvestrini, {\it ibid.} {\bf 477}, 321(1996).
\bibitem{Hagelin}J. S. Hagelin, S. Kelley, and T. Tanaka,
Nucl. Phys. {\bf B415}, 293(1994).
\bibitem{Raz}G. Raz, Phys. Rev. {\bf D66}, 037701(2002)
(hep-ph/0205310); Y. Nir and G. Raz, {\it ibid.} {\bf 66},
035007(2002).
\bibitem{Inami}T. Inami and C. S. Lim, Prog. Theor. Phys.
{\bf 65}(1981)297.
\bibitem{Grigjanis}R. Grigjanis, P. O'Donnell, M. Sutherland and H. Navelet,
Phys. Rep. {\bf 228}(1993)93; Phys. Lett. {\bf B213}(1988)355;
{\bf B286}(1992)413(E).
\bibitem{buras}G. Buchalla, A. J. Buras and M. Lautenbacher, Rev.
Mod. Phys. {\bf 68}(1996)1125.
\bibitem{Grinstein}B. Grinstein, R. Springer and M. Wise, Phys. Lett.
{\bf B202}(1988)138; Nucl. Phys. {\bf B339}(1990)269.
\bibitem{weinberg}S. Weinberg, Phys. Lett. {\bf B91}(1980)51.
\bibitem{Deshpande1}N. G. Deshpande and G. Eilam, Phys. Rev.
{\bf D26}(1982)2463.
\bibitem{Deshpande2}N. G. Deshpande and M. Nazerimonfared, Nucl. Phys.
{\bf B213}(1983)390.
\bibitem{Semertzidis}Y. K. Semertzidis {\it et al.},
hep-ph/0012087.
\end{thebibliography}
\end{document}